\documentclass[letterpaper,twocolumn,10pt]{article}
\usepackage{usenix,epsfig,endnotes}
\usepackage[normalem]{ulem}

\def\BibTeX{{\rm B\kern-.05em{\sc i\kern-.025em b}\kern-.08emT\kern-.1667em\lower.7ex\hbox{E}\kern-.125emX}}

\usepackage{xurl}
\usepackage{nicefrac}
\usepackage{siunitx}
\usepackage{array,framed}
\usepackage{booktabs}
\usepackage{dramatist}
\usepackage{adjustbox}
\usepackage{
  color,
  float,
  epsfig,
  wrapfig,
  graphics,
  graphicx,
  subcaption
}

\usepackage{textcomp}
\usepackage{setspace}
\usepackage{latexsym,fancyhdr,url}
\usepackage{enumerate}
\usepackage{algorithm2e}
\usepackage{algpseudocode}
\usepackage{graphics}
\usepackage{xparse} 
\usepackage{xspace}
\usepackage{multirow}
\usepackage{csvsimple}
\usepackage{balance}

\newcommand{\tool}{ScamChatBot\xspace}

\usepackage{pifont}
\newcommand{\circleone}{\ding{202}\xspace}
\newcommand{\circletwo}{\ding{203}\xspace}
\newcommand{\circlethree}{\ding{204}\xspace}
\newcommand{\circlefour}{\ding{205}\xspace}

\usepackage[framemethod=TikZ]{mdframed}
\definecolor{rubcolor}{HTML}{d1deeb}
\global\mdfdefinestyle{insightstyle}{%
backgroundcolor=rubcolor!3,
outerlinewidth=1pt,innerlinewidth=0pt,
outerlinecolor=rubcolor,roundcorner=5pt
}
\newmdenv[roundcorner=10pt, frametitle=Key Observations, linecolor=rubcolor]{insightbox}

\usepackage{
  tikz,
  pgfplots,
  pgfplotstable
}
\usepackage{hyperref}
\hypersetup{
    colorlinks=true,
    linkcolor=blue,
    filecolor=magenta,      
    urlcolor=blue,
    citecolor=blue,
    pdfpagemode=FullScreen,
}

\usetikzlibrary{
  shapes.geometric,
  arrows,
  external,
  pgfplots.groupplots,
  matrix
}

\pgfplotsset{compat=1.9}

\usepackage{mathtools,}

\DeclareMathAlphabet{\mathcal}{OMS}{cmsy}{m}{n}

\DeclareGraphicsExtensions{%
    .png,.PNG,%
    .pdf,.PDF,%
    .jpg,.mps,.jpeg,.jbig2,.jb2,.JPG,.JPEG,.JBIG2,.JB2}

\usepackage{xparse}
\newcommand{\bnm}{\begin{newmath}}
\newcommand{\enm}{\end{newmath}}

\newcommand{\bea}{\begin{eqnarray*}}%
\newcommand{\eea}{\end{eqnarray*}}%

\newcommand{\bne}{\begin{newequation}}
\newcommand{\ene}{\end{newequation}}

\newcommand{\bal}{\begin{newalign}}
\newcommand{\eal}{\end{newalign}}

\newenvironment{newalign}{\begin{align}%
\setlength{\abovedisplayskip}{4pt}%
\setlength{\belowdisplayskip}{4pt}%
\setlength{\abovedisplayshortskip}{6pt}%
\setlength{\belowdisplayshortskip}{6pt} }{\end{align}}

\newenvironment{newmath}{\begin{displaymath}%
\setlength{\abovedisplayskip}{4pt}%
\setlength{\belowdisplayskip}{4pt}%
\setlength{\abovedisplayshortskip}{6pt}%
\setlength{\belowdisplayshortskip}{6pt} }{\end{displaymath}}

\newenvironment{newequation}{\begin{equation}%
\setlength{\abovedisplayskip}{4pt}%
\setlength{\belowdisplayskip}{4pt}%
\setlength{\abovedisplayshortskip}{6pt}%
\setlength{\belowdisplayshortskip}{6pt} }{\end{equation}}

\newcounter{ctr}

\newcounter{mytable}
\def\mytable{\begin{centering}\refstepcounter{mytable}}
\def\endmytable{\end{centering}}

\newcounter{myfig}
\def\myfig{\begin{centering}\refstepcounter{myfig}}
\def\endmyfig{\end{centering}}

\newlength{\saveparindent}
\newlength{\saveparskip}
\newcommand{\E}{{\rm I\kern-.3em E}}

\renewcommand{\eqref}[1]{\mbox{Equation~(\ref{#1})}}

\def \part {part}

\renewcommand{\paragraph}[1]{\vspace*{6pt}\noindent\textbf{#1}\;}

\def \blackslug{\hbox{\hskip 1pt \vrule width 4pt height 8pt
    depth 1.5pt \hskip 1pt}}
\def \qed{\quad\blackslug\lower 8.5pt\null\par}

\newcounter{mynote}[section]

\newcommand\ignore[1]{}

\newcounter{rcnote}[section]

\newcounter{mrnote}[section]

\newcounter{fknote}[section]

\newcounter{anote}[section]

\DeclareMathSymbol{\mlq}{\mathord}{operators}{``}
\DeclareMathSymbol{\mrq}{\mathord}{operators}{`'}

\newcommand{\rhf}[2]{R_{f, \gamma}}

\DeclareDocumentCommand{\edist}{o o}{
  \ensuremath{
    \IfNoValueTF{#1}{{d}}{{\sf d}(#1,#2)}
  }
}

\newcommand{\olrk}[1]{\ifx\nursymbol#1\else\!\!\mskip4.5mu plus 0.5mu\left(\mskip0.5mu plus0.5mu #1\mskip1.5mu plus0.5mu \right)\fi}

\NewDocumentCommand{\indseq}{ O{1} O{r} }{{#1}\ldots {#2}}

\makeatother

\begin{document}

\title{ScamChatBot: An End-to-End Analysis of Fake Account Recovery \\ on Social Media via Chatbots}

\author{
{\rm Bhupendra Acharya\textsuperscript{\textsection}}\\
CISPA
\and
{\rm Dominik Sautter\textsuperscript{\textsection}}\\
Saarland University
\and
{\rm Muhammad Saad}\\
X (formerly Twitter)
\and
{\rm Thorsten Holz}\\
CISPA
}

\maketitle
\begingroup\renewcommand\thefootnote{\textsection}
\footnotetext{Both authors contributed equally to this research.}
\endgroup

\begin{abstract}
Social media platforms have become the hubs for various user interactions covering a wide range of needs, including technical support and services related to brands, products, or user accounts. Unfortunately, there has been a recent surge in scammers impersonating official services and providing fake technical support to users through these platforms. 
In this study, we focus on scammers engaging in such fake technical support to target users who are having problems recovering their accounts. More specifically, we focus on users encountering access problems with social media profiles (e.g., on platforms such as Facebook, Instagram, Gmail, and X) and cryptocurrency wallets. 
The main contribution of our work is the development of an automated system that interacts with scammers via a chatbot that mimics different personas. By initiating decoy interactions (e.g., through deceptive tweets), we have enticed scammers to interact with our system so that we can analyze their modus operandi. Our results show that scammers employ many social media profiles asking users to contact them via a few communication channels. Using a large language model (LLM), our chatbot had conversations with 450 scammers and provided valuable insights into their tactics and, most importantly, their payment profiles. This automated approach highlights how scammers use a variety of strategies, including role-playing, to trick victims into disclosing personal or financial information.
With this study, we lay the foundation for using automated chat-based interactions with scammers to detect and study fraudulent activities at scale in an automated way.
\end{abstract}

\section{Introduction}
\label{sec:intro}

In recent years, social media platforms have emerged as popular communication instruments for information exchange, opinion sharing, news gathering, and advertisement. Realizing the potential of social media becoming a one-stop shop that caters to all such needs of its users, popular brands have established their portfolios to engage with their customers and offer useful services such as technical support, dispute resolution, and discounts. As such, social media platforms have invariably paved the way for convenient interactions between businesses and customers for service requests and resolutions. A popular manifestation of this interaction is online technical support, where customers report issues about accessing their social media accounts or cryptocurrency wallets and expect a resolution service from their business counterparts. Users typically send these requests to the official profiles of the relevant business by making public posts and expecting a response from their customer service. However, there is an inherent risk in publicly posting such requests on social media, exposing users to scams~\cite{backlinkosocialmediausers, ftcsocialmediaattack}. Any user can impersonate an official business representative and offer fake technical support to the customer. In fact, this form of social engineering attack has significantly increased in recent years, and it is colloquially called {\em Technical Support Scam}~\cite{cnetsocialmediaattack, bluetickattack,perisci2018,miramirkhani2017NDSS,srinivasa2018WWW}.

This type of support scams can be launched under any pretext as long as a user is seeking online support and a scammer offering help by falsely claiming to be an expert. Across all scam typologies, the end goal is to gain the victim's trust and lure them into making a payment or surrendering sensitive information that can be monetized. Although technical support scams have been ongoing since the early days of social media, their prevalence has significantly increased in recent years~\cite{perisci2018,srinivasa2018WWW} especially targeting cryptocurrency users and users operating multiple social media accounts. This trend is not surprising, as cryptocurrencies have gained significant popularity due to the support of cryptocurrencies by some of the major online payment platforms, including PayPal. Moreover, with the increasing number of social media platforms, users with accounts on multiple platforms often experience account access issues due to lost passwords or account takeovers. 

In this paper, we focus on these two scam typologies and deploy techniques to uncover technical support scammers. 
Our proposed detection method is inspired by a recent work by Acharya et. al~\cite{acharyahoneytweet} where they conducted a deception study to bait scammers into revealing their identity. Their methodology involved requesting technical support through tweets, and upon receiving a response from the scammers, engaging with them through messages. Despite their methodology being highly effective, it had several limitations that could be improved to perform a more comprehensive study of the scam life cycle. In particular, the work~\cite{acharyahoneytweet} heavily relied on manually crafted messages during the scam engagement process, which reduced the scalability of their proposed system. 

\begin{figure}[tb]
\centering
\includegraphics[width=.47\textwidth]{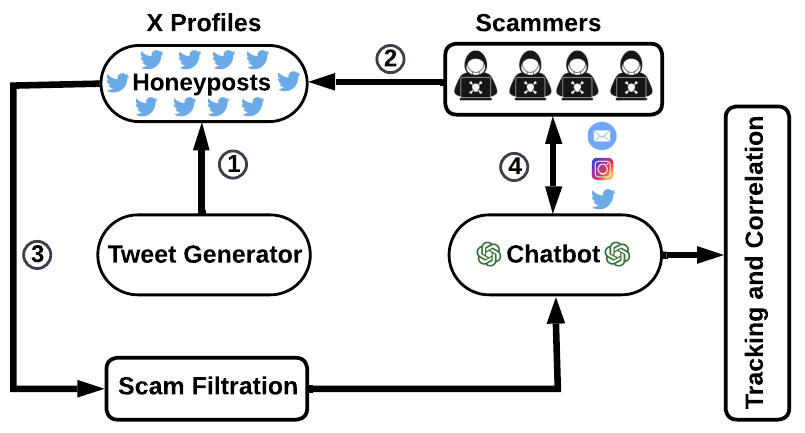}\hfill
\caption{\emph{ScamChatBot}: An overview of our system, which consists of three main components: i) \emph{Decoy scammers} which generate baiting tweets, ii) \emph{automated engagement} which performs a conversation with scammers via a chatbot, and iii) \emph{tracking and correlation}, which provides scammer's detail by tracking payment profiles and user social media profile data.}
\label{fig:sys_design}
\end{figure}

In this paper, we address the limitations and propose a more scalable and robust system that leverages more refined techniques to engage with scammers and gain deeper insights into the scam life cycle. With the rise of ML-based tools and Large Language Models (LLMs), we use the most advanced tools, ensuring efficiency and avoiding redundant efforts. More specifically, we use LLMs to craft human-style messages that are perceptually more believable and possess a higher propensity of trapping scammers. We introduce \emph{ScamChatBot}, a composition of automated techniques designed to trap technical support scammers targeting cryptocurrency and social media users. 
In~\autoref{fig:sys_design}, we provide a high-level workflow for \emph{ScamChatBot} which consists of three main components. The first component involves crafting baiting tweets (called \emph{honeyposts} \circleone) on X to lure scammers, posing as users experiencing account recovery problems across various account types. 
We regularly publish automated tweets via such deception accounts and monitor the real-time interactions of scammers with our tweets \circletwo. 
To ensure that we only interact with fraudsters, we filter out legitimate interactions \circlethree. 

The second component engages scammers who offer support through communication channels (often outside of X), initiating conversations, and seeking assistance via the chat module \circlefour.
To mimic more believable and human-style conversations, \emph{ScamChatBot} uses OpenAI's ChatGPT for automated conversations with scammers.
Our goal is to uncover their methods and especially information related to the payment profiles used in scams because the payment profile is crucial to understanding and disrupting fraud.

To this end, we have developed a chatbot that can pretend to be different personas to interact with scammers. 
We conducted conversations across three communication channels (Instagram, X, and email) because we found that these are the three channels commonly used by scammers.
Eventually, as the scam activity proceeds, scammers demand payments prior to offering technical support. We collect the payment methods and take appropriate actions to restrict the scam proliferation (i.e., sharing the collected information with payment providers). Novel to this work, our system exposes a worrying aspect of technical support scams. In prior works~\cite{acharyahoneytweet}, such scam campaigns were perceived to be operated by fraudsters who merely intended to pickpocket unsuspecting users. However, in our experiments, we discovered that scammers may also employ extortion techniques to threaten users with harm including personal information leakage. As a result, the impact of such scam campaigns extends far beyond financial losses, thus mandating the adoption of countermeasures and a necessity for more comprehensive works in this direction. 

In the measurement period between the last week of November 2023 and the end of March 2024, we generated about 19,000 honeyposts on our honey profiles that attracted over 11,700 scammers interacting with our system asking to contact them back via X direct messages, Instagram, and email. Scammer accounts with a large number of scammer profiles have been found to be operated by a single scammer~\cite{xiao2015detecting}. Through our analysis of scam network operation, we discovered that scammers utilize numerous social media profiles during interaction with honey profiles, and some scammers were found to employ as many as 71 social media profiles, prompting users to contact them via 1-2 communication channels. Among these scammers, we engaged in conversational dialogue with 450 scammers with our chatbot. By analyzing the timing of scammers' responses, we observed their promptness, typically replying on all three scamming channels within two hours of the conversation initiation and normally spending approximately one hour before realizing that the victim is unlikely to comply with any requested payment method. 

Through qualitative analysis of these conversations following a victim's engagement with a scam, we identified several techniques used by scammers. These include requesting live video or phone verification, soliciting private information, urging victims to pay for fake support, and potentially resorting to extortion or emotional manipulation if payment demands are not met. In addition, we measured ML-related text generation from scammers and found that 28.4\% of scammers are likely to use some form of automated text generation tools. In addition, analyzing the length of dialogue between the system and the scammer, we found that scammers on average reveal scamming payment profiles after seven interactive dialogues asking the system to pay for fake account recovery support, which highlights that an interaction is actually required to obtain relevant information. Furthermore, we collaborated with X, PayPal, and the cryptocurrency abuse database Chainabuse~\cite{trmslab}, sharing scamming PayPal addresses and cryptocurrency wallets that scammers asked the system to pay for fake support, and we received validation of scam activities. Our work thus provides a foundational model for the industry and security community to develop similar frameworks for identifying different categories of fraud on social media platforms.

\paragraph{Contributions} We make the following key contributions:
\begin{itemize}

    \item \textbf{Interactive Scam Detection.} We introduce \emph{ScamChatBot}, an interactive, scalable, and LLM-based framework capable of unveiling scammer life cycles and techniques through interactive dialogues. The framework can be generalized to decoy scammers of various scam categories across multiple social media platforms.

    \item \textbf{Detecting Fraudulent Payments.} Our system collects the scamming payment profiles of fraudsters in real time, which would otherwise remain inaccessible without engaging with scammers directly. In an empirical study, we analyze technical support scammers on social media.

\end{itemize}

\paragraph{Ethical Consideration and Data Disclosure} In keeping with the ethical guidelines required for a deception study, we obtained IRB approval from our institution. When managing data related to scammers, we strictly adhered to the GDPR and the EU Data Management Policies~\cite{eudatamanagement}. We conducted a deceptive study to ensure scammers remained unaware of being observed, preserving the authenticity of their responses and behaviors. Informing them or allowing withdrawal would have compromised the research, and retroactive debriefing potentially risks alerting other scammers and retribution attacks from scammers. Our study may have impacted scammers by risking the banning of their payment-related information after it was disclosed to PayPal and the cryptocurrency abuse database. However, this disclosure potentially prevents future attacks on other victims. Our tweet module was configured to adhere to ethical standards when tweeting. In total, we generated about 19,000 tweets across six different accounts over 101 days, a small number compared to the enormous volume of tweets on X (approximately 500 million per day). We posted innocuous tweets to attract scammers targeting users seeking help to recover their accounts. We made sure not to engage with official cryptocurrency wallet support teams or social media accounts. Thus, the sole objective of our system was to interact with the scammers by faking unsuccessful payment attempts after we were asked to pay for the fake account recovery in order to collect the fraudsters' payment profiles. Moreover, we communicated our findings to Meta, Google, X, ten cryptocurrency wallets, PayPal, and the cryptocurrency abuse database Chainabuse and received numerous positive responses. We worked closely with X, PayPal, and Chainabuse, confirming the scam validation and financial loss from these fraud profiles. 

We share our code at Github repository~\cite{scamchatbotGitRepo}. However, for data related to scammers, we refrain from publicly sharing to avoid potential retribution attacks and plan to provide it to interested researchers or academics upon request.
\section{System Design}
\label{sec:system_design}

In this section, we discuss the design of our framework \emph{ScamChatBot}, which creates unique tweets to bait scammers and performs on-demand chat initiation with scammers using a chatbot. 
As shown in Figure~\ref{fig:sys_design}, the system consists of three main elements to: (i) attract scammers via decoy interactions, (ii) an automated engagement mechanism powered by Large Language Models (LLMs) for conversations, and (iii) a tracking module that monitors and links the various interactions we have observed. 
In the following, we discuss the individual components in more detail.  

\subsection{Attracting Scammers via Decoys}

As a first step, we need to somehow get scammers to interact with us.
To do this, we create unique tweets (called \emph{honeyposts}) that mimic a request for technical support to attract scammers. The tweets ask for assistance with logging into social media accounts or support for cryptocurrency wallets. 
To generate honeyposts, we have designed tweets based on two main criteria: i) adherence to the length limits set by X posts~\cite{twitterlengthmax} and ii) addressing users experiencing account access issues. These tweets are tailored to be posted every 15 minutes on X in English. To compose a tweet, we use logic with a three-sentence structure. The first sentence usually contains introductory phrases such as ``Good morning'', ``What happened'', or ``I can't believe it''. The second sentence, which is often coupled with the first, is aimed at users who are having problems with their accounts. Finally, the third sentence asks for urgent help, with or without a hashtag. These hashtags are randomly generated based on account-related keywords.

We utilized six X profiles, which we referred to as \emph{honey profiles}, each dedicated to posting unique content for these accounts. Three profiles focused on providing social account recovery support, while the remaining three focused on support for wallet-related recovery as follows:

\begin{itemize}
\item \textbf{Social Media Account Honeyposts}: For social media tweets, we selected five popular services including Gmail, Instagram, Youtube, X, and Facebook based on their popularity in Google searches~\cite{stat2024popularsocialmedia, forbes2024popularsocialmedia}. We then generated tweets in which the account pretends to be a user of those services experiencing login issues.  
    
\item \textbf{Cryptocurrency Honeyposts}: 
For cryptocurrency-related tweets, we selected ten popular wallets reported in a prior study performed in a similar setup~\cite{acharyahoneytweet}. The wallets included \emph{Badger}, \emph{Binance}, \emph{BitPay}, \emph{Coinbase}, \emph{Exodus}, \emph{Free}, \emph{Ledger}, \emph{MetaMask}, \emph{Trezor}, and \emph{Trust Wallet}. Similar to the context of social media login issues, we generated tweets in which the account reports login issues with their cryptocurrency wallets and requests account recovery. 
\end{itemize}

\subsection{Automated Engagement with Chatbots} 
In this section, we provide an overview of the filtration techniques implemented to ensure automated chat conversations are performed exclusively with identified scammer communication channels, as well as the necessary chat configuration and account setup for engaging with scammers. Below we provide such details.

\subsubsection{Engagement Pre-processing Filtration} The tweets posted by \tool prompt scammers to interact with them through \emph{replies} and \emph{quoted tweets}. In those interactions, scammers typically provide their contact details, such as email or social handles, asking for further interactions. As a follow-up to the scammer's communication channel, we implemented an engagement mechanism that initiates an automated chat dialogue conversation with the scammer's social media profile. To ensure that we do not interact with benign accounts in an automated way, we apply various filtration techniques before initiating a chat interaction, such as excluding verified X accounts, official social media platforms such as X, Google, cryptocurrency wallet services, and benign accounts. We provide further details on filtration techniques below. 

\begin{itemize}
\item \textbf{Exclude Verified Accounts.} The engagement module removes all verified accounts that respond to the tweets. Typically, verified accounts are associated with official representatives of a platform, and they may respond in a sincere attempt to offer technical support. We therefore remove them from our study. We acknowledge that recent changes in X's verification policy can lead to scammers acquiring fake verification for their profiles~\cite{misuseverifiedontwitter, twitterbrandimpersonatorverified,cryptoscamsbyverifiedattack}. However, we preferred a more conservative approach in an attempt to remove potential false positives even at the risk of ignoring true negatives.

\item \textbf{Exclude official social media services.} In the current setup, the engagement module also excludes the top 20 social media services along with the top 100 wallet services to ensure that our chat module does not initiate any conversation with official support services. We collected  20 popular social media services ~\cite{stat2024popularsocialmedia,forbes2024popularsocialmedia} and official cryptocurrency wallet services~\cite{popularcryptowalletscoincarp,popularcryptowalletstrustradius} based on social media popularity search. 

\item \textbf{Benign Account Filtration.} For benign account filtration, we faithfully replicated the approach in~\cite{acharyahoneytweet} where the authors identified two key characteristics of scam accounts that were distinctly different from benign accounts. During the engagement process, scam accounts provide new communication channels (often outside of X) to continue the conversation and complete the scam activity. Moreover, scammers also pose as official representatives of the platform to gain the user's trust. In contrast, there are also benign accounts that interact with users, often providing advice or expressing sorrow for the user's despair. Those accounts do not interact with users beyond the initial interaction, thus displaying no indication of fraudulent intentions. We label such accounts as benign and exclude them from our study. 
\end{itemize}

Additionally, we provide a manual qualitative analysis in Section~\ref{sec:manual_analysis:filtration}, which shows that the filtration techniques work well in practice.

\subsubsection{Scammer Engagement} After the account filtration process, the engagement module creates personas using a chatbot based on an LLM such as ChatGPT to interact with the scammers on the platform channels they provided. For each communication channel, including Instagram, Gmail, or even X, we generate unique personas with the purpose of (i) understanding the scam activity, and (ii) extracting payment information provided by scammers. Therefore, the ChatGPT-configured persona mimics a human-like behavior while conversing with scammers and leverages context awareness and organic articulation approaches to gain the scammer's confidence. In every new conversation, we ensure that there is no LLM-related text present that could be picked up by the scammer. Moreover, before our live experiments, we also conducted manual conversations with \emph{ScamChatBot} to ensure it was fully capable of performing organic conversations and collecting the desired information, especially related to payment profiles. 

\paragraph{Dialogue Generation and Direct Messaging} \emph{Dialogue} refers to the conversation between the scammer and our system aimed at resolving account-related issues. To facilitate dialogue communication, we selected communication channels based on fraudsters' popularity asking our system to contact for fake support. These platforms consist of three channels: Email, X, and Instagram direct messages. In total, we create 15 distinct accounts (i.e., 5 per platform: X, Instagram, and Email). We allocate four accounts for each channel for the personas involving cryptocurrency and social media account recovery, while the remaining three are designated for language-based experimental personas. For each communication channel, we integrate the respective \emph{APIs}~\cite{instagramAPI, twitterAPI, protonAPI} to receive real-time updates when the scammer interacts with our system. Upon receiving messages from the scammer, we aggregate all previous interactions between the system and the scammer and generate a response based on the personas. Subsequently, we deliver the generated message to the scammer via each API as our response. In our direct messaging interactions with scammers, we avoid using the accounts designated for tweets. We separate our direct messaging accounts from our X platform tweet post accounts to effectively communicate with the scammers without suspicion. 

We ended our conversations in two cases: (i) after the scammers provided multiple payment profiles (e.g., cryptocurrency addresses and PayPal email addresses) and (ii) when the scammers stopped communicating with us. In the first scenario, we concluded by stating that we would explore further alternatives (after receiving all payment profiles from the fraudsters). In the second scenario, we made at least one attempt to reconnect and asked the scammer if they would like to continue the chat in order to receive further assistance.

In Section~\ref{sec:manual_analysis:suspicion}, we present the results of a manual qualitative analysis of our system's ability to maintain context continuity and discussion on limitations in~Section~\ref{sec:discussion}. We empirically find that the approach is able to effectively identify payment methods from scammers. 

\paragraph{Persona Types and Configuration} ChatGPT requires a specified persona configuration to delineate its role in direct message interactions with scammers. We establish two primary personas for ChatGPT's configuration. The initial persona involves a user having issues with a cryptocurrency wallet, particularly tailored for newcomers to cryptocurrency or crypto investment who find themselves unable to access their wallet. The second persona simulates a user encountering difficulties accessing a social media platform such as X, Instagram, Facebook, or Gmail. In this scenario, the user communicates their inability to access their account and seeks assistance in recovering it. Furthermore, we conduct an experiment that expands upon these personas by introducing a user who lacks proficiency in English and desires to switch to other languages. In this context, the system initially prompts the scammer to determine if they are conversant in languages other than English and endeavors to communicate in the languages agreed upon during the dialogue.

Additionally, the persona includes additional rules in text generation. During the conversation, as the scammer asks questions about the account-related issues, the persona is asked to create a fake reason for having an access problem. Scammers often ask for some form of payment to resolve such an account issue via various payment methods. While doing so, scammers create a fake reason such as needing to buy software to recover the account or sending money to re-enable the account-blocked workflow. The persona set for ChatGPT also tries to ask for multiple payment methods so that the scammer can reveal as many payment methods as possible. For this, we take a deceptive approach by falsely claiming failure in the first payment method provided by the scammer. We repeat the process until the scammer runs out of options to share with us, at which point the engagement module ends the conversation. We provide additional detail on the configuration of chat persona and sample conversation between our system \emph{ScamChatBot} and a \emph{Scammer} in Appendix~\ref{sec:chat_configuration}.

\subsection{Qualitative Evaluation of Dataset}
\label{sec:manual_analysis}
In this section, we provide a manual qualitative evaluation of our dataset across several categories. To assess the effectiveness of our system conversations with scammers, we randomly selected 15 conversations each from Instagram, email, and X platform, evaluating a total of 45 interactions. Below, we provide details of our evaluation findings.

\subsubsection{Legitimacy of Data Filtration and Methodology Accuracy} 
\label{sec:manual_analysis:filtration}
In this category of qualitative analysis, our focus was to identify the accuracy of our methodology. Specifically, we looked for instances where interactions occurred between our system and both malicious and legitimate accounts. Out of 45 conversations from three platforms, we found no instances where the system mistakenly engaged with legitimate users. Additionally, we observed that none of the conversations involved official support accounts.

\subsubsection{Repetition in Conversation/Suspicion}
\label{sec:manual_analysis:suspicion}
In this category of qualitative analysis, our focus was on observing our system's ability to maintain context continuity. We assessed the quality of conversations, particularly those involving repetitive exchanges or discussions where scammers attempted to determine if the system was a bot. We identified 3/45 conversations in which scammers noted repetitive apologies for not being able to send screenshots due to technical problems and perceived a lack of seriousness in taking action. In all three cases, scammers expressed suspicion, stating either that the system was likely a robot or that it was wasting time without genuine intent to resolve the issue. While the scammers became suspicious that they were interacting with a bot, they still disclosed their payment profiles. This demonstrates that our system was effective in identifying fraudulent payment profiles. Such identification can enable proactive measures to block fraudulent payments and prevent further scamming of potential victims.

\subsubsection{Limitation to External Channel Communication}
\label{sec:manual_analysis:limitation}
In this category of qualitative analysis, our system faced limitations in cases where scammers redirected communication to an external channel. Specifically, in three instances, scammers on X requested to connect via direct message on another X account. Since our system was not configured to interact through newly introduced channels during the conversation, these interactions did not proceed in-depth. We identified three accounts—two from X and one from email—that attempted this redirection. In one case, the system politely ended the conversation, while in the other two, it falsely claimed to have already contacted the referenced person and received no response. Additionally, in one email interaction, the scammer provided a template asking for cryptocurrency addresses and a passphrase. In all of these three instances, our system had a limitation in further communicating with the scammer-provided channel.

\subsection{Tracking and Correlation} 
The third major component of \emph{ScamChatBot} is the tracking module, which conducts analysis on data collected by the engagement module. For the purpose of this study, the key features we analyze include social media profile metadata analysis, conversation text created by our system and responses received from scammers, payment methods shared, and the end-to-end scam lifecycle. 

\medskip
In the following, we present our main finding.
In Section~\ref{sec:delineate}, we provide details on the interaction with decoy profiles and the automated chat module, while we present the timing and operation analysis of scammers in Section~\ref{sec:scammer_operations_patterns}. Section~\ref{sec:qualitative_study} focuses on a qualitative study of the interaction between our system and scammers, and we focus on the tracking of financial losses in Section~\ref{sec:financial_loss_metrics}. 
\begin{figure*}[tb]

\centering
\includegraphics[width=.33\textwidth]{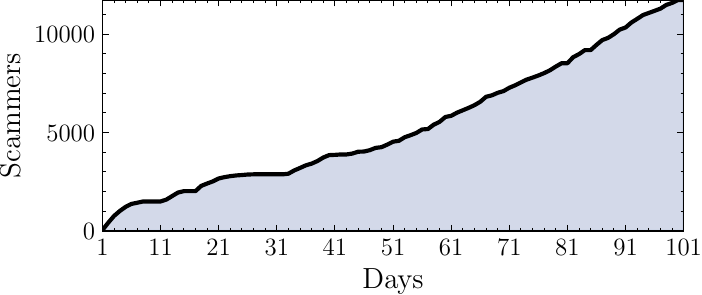}\hfill
\includegraphics[width=.32\textwidth]{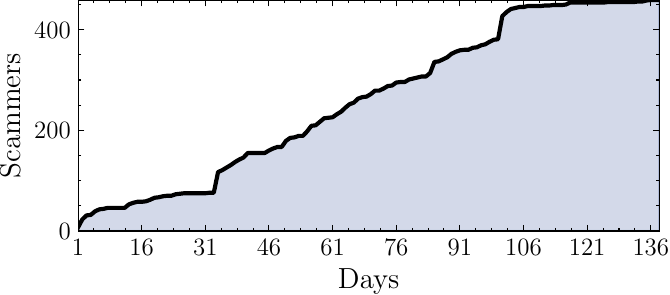}\hfill
\includegraphics[width=.32\textwidth]{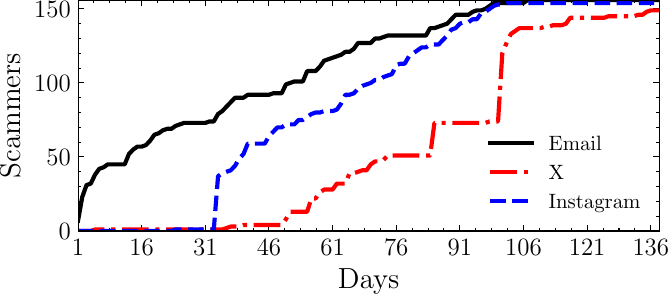}

\caption{Interactions of scammers with our decoy profiles and our chatbot. The left graph shows the cumulative sum of scammers interacting with our honeyposts over the experiment duration. The middle graph shows the daily cumulative sum of scammers engaged via conversation with our system \emph{ScamChatBot}. The right graph shows the cumulative sum of scammers based on various communication channel engagements with \emph{ScamChatBot}. }
\label{fig:scammer_interaction_with_scam_chat_bot}

\end{figure*}

\section{Delineating Scammers Interaction}
\label{sec:delineate}
We start by discussing results regarding interactions between scammers and our system.

\subsection{Interaction with Decoy Profiles} Between mid-November 2023 and mid-February 2024, the tweet generator posted unique tweets for 101 days. During this period, we posted tailored tweets aimed at luring scammers who target users seeking assistance with account recovery.  
Figure~\ref{fig:scammer_interaction_with_scam_chat_bot}(a) (left) shows an overview of scammers interacting with our honeyposts and we present the overall tweet posts interaction results of our honey profiles in Table~\ref{table:overall_tweet_module_interaction}. Out of 19,001 tweet posts, the module posted 32.74\% (6,221/19,001) tweets related to social media recovery of five different social media accounts and 67.3\% (12,780/190,001) tweets related to cryptocurrency wallet recovery of ten different wallets. Out of 6,782 tweets that received an interaction, the interaction percentage of social media-related account recovery was 10.8\% (674/6,221) compared to a higher percentage of cryptocurrency wallets at around 47.8\% (6,106/12,780). Overall, among all posted tweets, 35.7\% (6,782/19,001) received interactions in the form of post comments from 11,769 potential scammers. Scam accounts replied with 22,285 distinct messages to all honey profiles. Among them, 90.0\% (20,046/22,285) tweets were sent in response to cryptocurrency-related accounts and tweets.  

In our initial analysis, we hence find that cryptocurrency users are more commonly targeted by scammers, likely due to a higher financial incentive. We exclude scammers performing non-text interactions, such as retweets, likes, follows, bookmarks, and impressions. Further details on non-text interactions are provided in the Appendix~\ref{sec:non_text_interaction}.

\begin{table}[t]
\centering
\setlength\tabcolsep{5pt} 
\renewcommand{\arraystretch}{1.7}
\scriptsize
\begin{tabular}{lrrrrr}
\toprule
\bf{Account} & \bf{Tweets} & \bf{Interacted} & \bf{Distinct} & \bf{Total} & \bf{Interacted}  \\
\bf{Recovery} & \bf{Posted} & \bf{Tweets} & \bf{Text} & \bf{Text} & \bf{Scammers}  \\
\midrule
Crypto Wallet & 12780 & 6106 & 20046 & 29652 & 10647 \\
Social Media & 6221 & 674 & 2239 & 2691 & 1581 \\
\midrule
\bf{All} & \bf{19001} & \bf{6782} & \bf{22285} & \bf{32343}& \bf{11769} \\
\bottomrule
\end{tabular}
\caption{Summary of account recovery chat module based on tweets interacted by scammers from two account recovery contexts: cryptocurrency wallet and social media-based account recovery.}
\label{table:overall_tweet_module_interaction} 
\end{table}

\subsection{Interaction with Automated Chat Module} From the last week of November 2023 to the end of March 2024, our chat module engaged in communication with scammers via communication channels posted through tweet posts. We continued to engage with scammers for 6 weeks even after ending our tweet module to accommodate those scammers who persisted in engaging with our system. Figure~\ref{fig:scammer_interaction_with_scam_chat_bot}(a) (left) displays the interaction of scammers with our honeyposts, whereas 
Figure~\ref{fig:scammer_interaction_with_scam_chat_bot}(b) (middle) displays the number of scammers interacting with our system via the chat module.
Following thorough filtration, as outlined in Section~\ref{sec:system_design}, our system initiated automated chats via three communication channels: email, X, and Instagram direct messaging. 
Figure~\ref{fig:scammer_interaction_with_scam_chat_bot}(c) (right), we present the number of interactions via each communication channel.
Scammer's preference for email shows that not all scammers operate within the instant messaging model. In the following, we dive deeper into the modus operandi of scammers and provide an overview of their interactions.

\paragraph{Scam Lifecycle} For both account recovery contexts, scammers typically follow a consistent pattern: initially, scammers begin by creating a verification process for the victim's account information. Once some form of verification is established, the scammer fabricates a false pretext and requests payment in various forms. Following this initial stage of verification, the scammer often instructs the potential victim to conduct an internal audit and then return with the details. Finally, the scammer suggests that after the payment is made, it may take some time for the account to be fully recovered due to essential technical actions required for recovery. Below, we provide further insights into detailed operational cycles performed by scammers.

\paragraph{Victim Verification Procedure} Initially, the scammer requests information such as the victim's account information (e.g., wallet address or social media account), the timing of the issue occurrence, and any actions such as withdrawals or login actions performed lately. This initial conversation is followed by additional details to establish identities, such as location, email, name, phone number, and a video recording of the issue with the wallet or account. Oftentimes,  scammers try to initiate communication via WhatsApp or Facebook to further verify the victim's authenticity. During this verification phase, scammers frequently request screenshots of the account, and in some instances, they may also ask for secret phrases.

\paragraph{Advance Fee} After the victim verification process, scammers request an advance fee to initiate the recovery process. The advance fee is often requested under the pretext of purchasing software (e.g., \emph{Zeus Software} or \emph{Trojan Horse}) which helps with the recovery process. In some cases, scammers also request seed phrases for wallets, which is one of the easiest methods to steal money. Seed phrases can be easily converted into private keys, which can then be used to transfer cryptocurrency assets to a new address. 

\paragraph{Payment Method Instructions} Scammers request payment for their services through various methods. We observed different kinds of payout methods, e.g., Amazon or Apple gift cards, cryptocurrencies such as Bitcoin and Ethereum, or other online payment methods such as PayPal, Venmo, and CashApp. Alternatively, the scammer might direct the victim to an external link for payment. In the case of PayPal, scammers are found to instruct the victim to use the ``send to family and friends'' option. Upon consulting PayPal, we learned that payments through this option are not eligible for refund or dispute. Therefore, once the scam activity is completed, the victim cannot dispute the charge with PayPal. Finally, when the payment is completed, scammers also request a payment screenshot for confirmation. 

\paragraph{Scammer's Reaction to Failed Payment Attempt} As mentioned earlier, we used failed payment excuses to bait scammers into revealing all their payment methods. We noticed that scammers expressed resentment or frustration when we mentioned payment failures to them. Occasionally, they also used profane language and threatened to cause personal harm. Although we were using fake personas while interacting with scammers, it is indeed concerning since real users might have experienced such unwarranted abuse in the form of threats to them and their families. It is therefore pertinent to curtail this form of technical support scam, as it can also transition into extortion and abuse. 

\paragraph{Results Overview} In Table~\ref{table:overall_messages_interacted}, we provide an overview of our conversations. Out of 11,769 scammers that interacted with our honey profiles, our system interacted with 450 scammers across three communication channels: X (149), Instagram (154), and email (156) excluding accounts that opted out of direct messaging (994) and scammers asking external communication channels such as Phone, including WhatsApp (131), Telegram (121), account restore URLs (199), and Google Forms (424). In the email conversations, out of 640 scammers to which we sent messages, 24.37\% (156/640) scammers replied with 681 total and 666 distinct replies. Among the three communication channels, Instagram was found to have the highest engagement replies (6,747) from replies with a total message of 73.2\% (13,668/18,678) collected by our system. Of the total of 18,678  messages recorded by the system, 46.7\% (8,721/18,678) were written by scammers.

\begin{table}
\centering
\caption{Breakdown of \emph{ScamChatBot} engagement with scammers. Within this table, we delineate the exchange of messages between our system and scammers across email, X, and Instagram platforms.}
\scalebox{0.91}{
\renewcommand{\arraystretch}{1.3}
\setlength\tabcolsep{3pt} 
\begin{tabular}{lrrrr}
\toprule
\bf{Channels} & \bf{Msg. Sent} & \bf{Msg. Replied} & \bf{Scammers} & \bf{Total}\\
\bf{Interaction} & \bf{Dist/All} & \bf{Dist/All} & \bf{Replied/Sent} & \bf{Msg}\\
\midrule
Email & 1325/1327 & 666/681 & 156/640 & 2008 \\
X & 1303/1709 & 1200/1293 & 149/276 & 3002 \\
Instagram & 4168/6921 & 3286/6747 & 154/171 & 13668 \\
\bf{All} & \bf{6788/9957} & \bf{5152/8721} & \bf{450/1087} & \bf{18678} \\
\bottomrule
\end{tabular} }
\label{table:overall_messages_interacted}
\end{table}
\section{Investigation into Scammer Tactics}
\label{sec:scammer_operations_patterns}
In this section, we analyze several intricacies, such as analyzing the timing of scammers' engagement with our system based on how quickly they respond, examining operations based on the day of the week, exploring the potential use of ML-generated messages, investigating organized networks of scammers that may utilize multiple social accounts as part of coordinated abuse, and dialogue necessities as part of revealing scamming payment profile. 

\subsection{Timing Analysis}
We start by presenting a timing analysis to showcase two key aspects of scam engagement methods. In the first analysis, we explore the time taken by fraudsters to respond to us after our initial engagements. In other words, once the scammer informed us of the method to connect with them and our system sent a help request, how long did the scammers take to respond to us. As a result, we could assess the communication channels being actively monitored by scammers, indicated by their swift response time.  In the second analysis, we study the total duration of our engagement with the scammer where they persistently tried to trap us in hopes of obtaining any financial benefit. Our second analysis revealed the persistence shown by scammers in engaging with their victims. 

\begin{table}
\vspace{2ex}
\centering
\caption{Breakdown of scammer's first response with our system \emph{ScamChatBot}. The data represents how quickly scammers respond to our system upon asking for account recovery support.} 
\scalebox{0.89}{
\renewcommand{\arraystretch}{1.07}
\setlength\tabcolsep{3pt} 
\begin{tabular}{lrrrr}
\toprule
\bf{Channels} & \bf{Median} & \bf{90 } & \bf{Response} & \bf{Total}\\
\bf{Interaction} & \bf{Response} & \bf{Pct} & \bf{$<$ 1000 ms} & \bf{Scammers}\\
\midrule
Email & 0:33:09 & 11:48:45 & 48 & 156 \\
X & 2d 6:16:36 & 17d 10:50:42 & 13 & 149 \\
Instagram &  0:57:18 & 8d 16:05:36 & 9 & 154 \\
\bf{All} & \bf{1:48:54} & \bf{9d 15:48:43} & \bf{70} & 450\\
\bottomrule
\end{tabular} }
\label{table:scammers_first_response_timing}
\end{table}

\subsubsection{Initial Response Time} Our analysis of the initial response time revealed that scammers are more likely to respond faster via email compared to those on X and Instagram (see Table~\ref{table:scammers_first_response_timing}). The median response time for email was 33 minutes and 9 seconds which was considerably faster than X, where the median response time was over 2 days. During our analysis, we identified that some communication channels indicated scripted responses by replying to our requests within 1 second. Upon closer inspection, we found that 30.8\% (48/156) of initial email responses were automated replies containing additional information or directing users to release secret keys via external sites. Similarly, for X and Instagram, we observed automated setups for first-time accounts, accounting for 8.7\% (13/149) and 5.8\% (9/154), respectively. In total, we identified 15.6\% (70/450) of communication channels employing some form of initial auto-script to entice victims into divulging private information. Overall, scammers' response times exhibited a median of 1 hour and 48 minutes, with a 90th percentile response time of 9 days.

\begin{table}
\centering
\caption{Breakdown of total time scammers performing dialogue - The difference between the first and last responses from scammers interacting with our system \emph{ScamChatBot}.} 
\scalebox{0.75}{
\renewcommand{\arraystretch}{1.3}
\setlength\tabcolsep{3pt} 
\begin{tabular}{lrrrrr}
\toprule
\bf{Channels} & \bf{Median} & \bf{90 } & \bf{Response} & \bf{One Time} &\bf{Total}\\
\bf{Interaction} & \bf{Time Spent} & \bf{Pct} & \bf{$<$ 1000 ms} & \bf{Reply} & \bf{Scammers}\\
\midrule
Email & 0:18:00 & 5:57:18 & 52 & 52 & 156\\
X & 1d 2:03:56 & 13d 15:32:39 & 15 & 15 & 149 \\
Instagram &  0:57:03 & 5d 16:30:46 & 4 & 4 & 154 \\
\bf{All} & \bf{0:57:30} & \bf{6d 22:30:37} & \bf{71} & \bf{71} & 450\\
\bottomrule
\end{tabular} }
\label{table:scammers_total_time_spent_with_our_system}
\end{table}

\subsubsection{Total Time Spent Analysis} In Table~\ref{table:scammers_total_time_spent_with_our_system}, we present the duration scammers spent engaging with our system before discontinuing further conversation. Scammers communicating via email channels were found to spend a total of 18 minutes in conversation, which was notably lower compared to those communicating through X, spending about 1 day and 2 hours, and Instagram, spending about 57 minutes. 52 out of 156 emails received from scammers resulted in only one-time replies with further instructions, while for X and Instagram, accounted 10.06\% (15/149) and 2.59\% (4/154), respectively. Overall, scammers spent an average of 57 minutes and 30 seconds, with the 90th percentile spending approximately 6 days and 22 hours before discontinuing further communication. We observed that scammers who are active for more than a day tend to repeatedly ask if the victim needs additional assistance with account recovery and often mention that they will resume account recovery after successful payment. 

\begin{figure}[tb]
\centering
\includegraphics[width=0.45\textwidth]{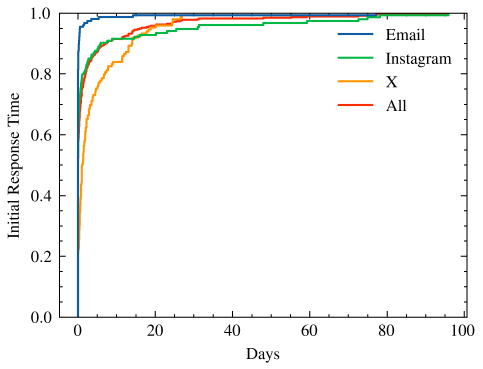} 
\caption{Number of days scammers spent time with our system - In this graph we show the difference between the first response and last response from scammers interacting with our system.}
\label{fig:number_of_days_scammer_spent_time_with_our_system}
\end{figure}

In Figure~\ref{fig:number_of_days_scammer_spent_time_with_our_system}, we show the total number of days spent by scammer interacting with our system: more than 80\% of the scammers would leave interacting after the first-day interaction with our system, and the remaining 20\% would follow up to continue the conversation from the last point. 

\subsection{Operation Analysis}
In our analysis of operations, we (i) dive deep into scammers' weekly operation pattern, (ii) explore how scammers create numerous social media profiles and use a limited number of communication channels to prompt victims to initiate contact, and (iii) as ML and advanced text-generation tools become more prevalent, we assess the likelihood of scammers utilizing such technology in their interactions with our system. 

\subsubsection{Week Period Scammer Communication Analysis} 
\begin{figure}[tb]
\centering
\includegraphics[width=.4\textwidth]{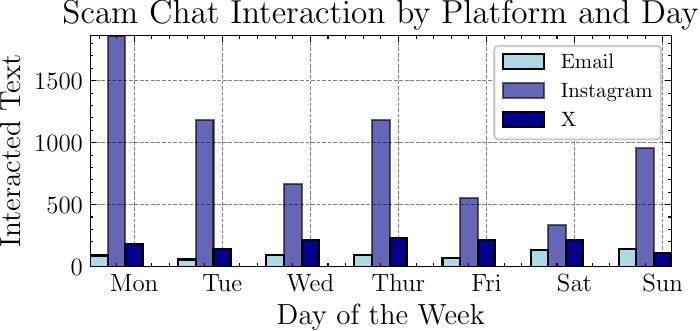}\hfill
\caption{Days of the week chat dialogue interactions - the graph shows text (chat dialogue) scammers performed via three different communication channels with our system \emph{ScamChatBot} throughout the days of the week. }
\label{fig:scammer_week_days_chat_dialogue_interaction}
\end{figure}

Our analysis of the days of the week interaction by scammers with our framework showed distinct patterns. The number of text interactions via chat dialogue shows that interaction patterns vary across different communication platforms (see Figure~\ref{fig:scammer_week_days_chat_dialogue_interaction}). For instance, scammers using Instagram as a communication channel for chat dialogue tend to be more active on alternate days, with the lowest activity observed on Saturdays. Conversely, scammers utilizing email exhibit increased activity on weekends compared to other weekdays, with alternating lows on Tuesdays and Fridays. Similarly, on X, scammers maintain consistent chat activity from Wednesday to Saturday, with reduced communication observed on Tuesdays and Sundays. 

\begin{figure*}[tb]
\centering
\includegraphics[width=0.328\textwidth]{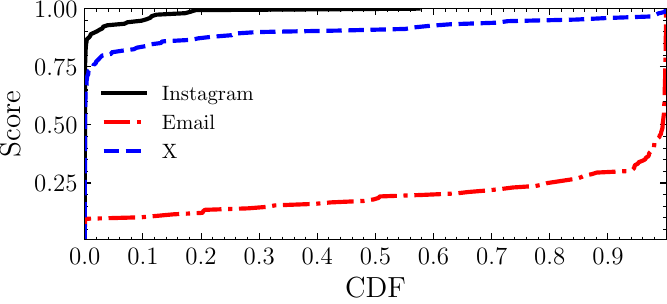} 
\includegraphics[width=0.328\textwidth]{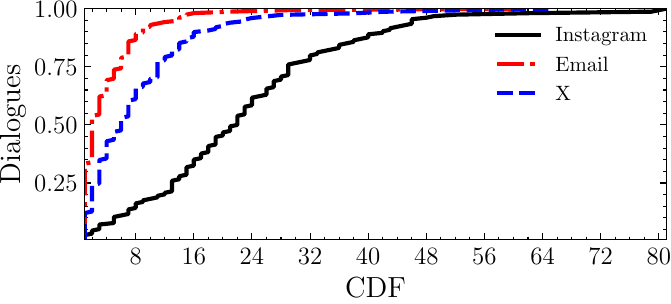} 
\includegraphics[width=0.328\textwidth]{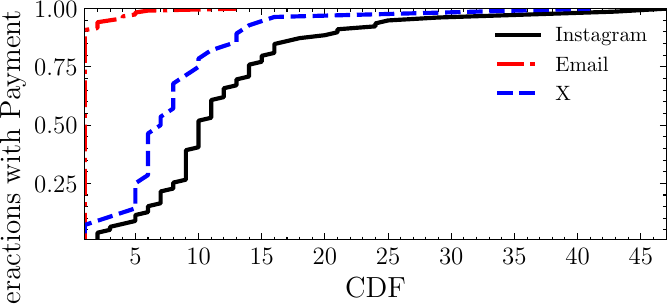} 

\caption{Graph on analysis of dialogue - In the left graph (a) we show the ML text score from each social media dialogue from scammers. In the middle graph (b), we show the total dialogue length between our system and scammers from each platform, while on the right side (c), we illustrate the total count of dialogue interactions needed from three platforms before the scammer discloses the payment method to request payment from the system.
}
\label{fig:text_dialogue}
\end{figure*}

\begin{table*}
\centering
\vspace{2ex}
\scalebox{1}{
\renewcommand{\arraystretch}{1.3}
\setlength\tabcolsep{3pt} 
\begin{tabular}{lrrrrrrrrr}
\toprule
\bf{Channels} & \bf{Mean} & \bf{Median} & \bf{Min} & \bf{Max} &\bf{Score $>=$50} &\bf{Score $>=$75} &\bf{Score $>=$90} &\bf{Total}\\
\midrule
Email &  0.80820 & 0.99419 & 0 & 0.99996 & 128 & 120 & 110 & 156\\
X & 0.09651 & 0.00013 & 0 & 0.99986 & 14 & 8 & 7 & 149\\
Instagram & 0.01323 & 1.57e-05 & 0 & 0.57432 & 1 & 0 & 0 & 154\\
\bf{All} & 0.30095 & 0.00092 & 0 & 0.99996 & 143 & 128 & 117 & 450\\
\bottomrule
\end{tabular} }
\caption{Breakdown of AI text score from each communication channel. In this table, we present the AI text score for each scammer's communicated text and the respective score. Scammers communicating with emails are found to utilize some form of AI text generation compared to other social media platforms. } 
\label{table:ai_text_score_table}
\end{table*}

\subsubsection{Scam Network Operation} In this section, we perform a collaborative network analysis, centering on (i) the creation of scammer profiles, which includes examining names, descriptions, followers, and following; (ii) the utilization of shared communication channels by scammers such as instances of scammers sharing a common communication channel through tweet posts, urging potential victims to contact them for technical support. 

\paragraph{Shared Profile Information} We perform an analysis of scammer profile details, including names, descriptions, followers, and following. Our investigation into scam networks revealed that 24.5\% (2,877/11,764) scammers shared a common name, 33.6\% (3,949/11,764) shared a common description, and 5.2\% (610/11,764) scammers were found to have multiple common followers in their profiles. Among these 610 shared followers, we detected 503 instances of mutual following to more than one account. This leads us to suspect that these profiles represent a cluster of organized scamming operations aimed at executing targeted scamming campaigns.

\paragraph{Shared Communication Channels} Our analysis of clustering scamming profiles by communication channels indicates that 42.6\% (5,012/11,769) of the profiles engaged in scamming activities were observed to use more than one communication channel. Within this subset of 5,012 scamming profiles, the highest number of communication channels employed by a single profile was 71, while the median value for other scamming profiles was 3 communication channels. Additionally, we found that 16.5\% (1,939/11,769) of the scamming accounts used all three communication channels, prompting users to contact them.

\subsubsection{Potential Operation of ML ChatBot}
Next, we perform an analysis of conversations originating from scammers, possibly generated using ML tools. Initially, we investigated ML text detection tools available in the market~\cite{forbestbestAIDetector}. To validate the authenticity of the text, we conducted an exploratory free trial of the top five services listed~\cite{forbestbestAIDetector}, comparing inputs of both ML-generated and non-ML-generated content. Among these services, we selected \emph{Sapling API}~\cite{saplingAPI} to further assess the likelihood of ML-generated text across all three communication channels. 
The APIs provided by Sapling for detecting ML content evaluated the aggregated scoring of the text from each scammer as the calculation of the mean score. These scores are derived from specialized models: one is trained to classify extensive portions of text, while the other is a perplexity-based model~\cite{saplingscoring}. In Figure~\ref{fig:text_dialogue} and Table~\ref{table:ai_text_score_table}, we summarize the score 0 to 1, where 0 represents the least likelihood and 1 represents the highest probability score of utilizing ML tools in the generation.

Our observations on ML-related text generation on scammers showed 28.4\% (128/450) of scammers with a probability of 50 or more found to be likely to use some form of ML text generation tool in text creation. Similarly, on X and Instagram, a probability score of greater than 50, was found to be 9.4\% (14/149) and 0.6\% (1/154) scammers, respectively. In a note to scammers with a probability of 75 or greater was found to be 28.4\% (128/450). Overall, from all three communication channels, 26.0\% (117/450) with a probability score of 90 or greater are likely to use some form of ML text generation tool in communicating with our system. Furthermore, in Figure~\ref{fig:text_dialogue}(a) we show the score of each communication channel.

\subsection{Analysis of Dialogue Length and Scamming Payment Information Disclosure} We examine the duration of interactions within our system and with scammers, focusing on the total dialogue length and the minimum dialogue required before payment details are disclosed or requested. This section provides a comprehensive overview of both aspects. 

\paragraph{Dialogue Length} Regarding the total dialogue length, the median dialogue length for the system was 8, while for scammers it was 7. The maximum dialogue length observed was 135 from the system and 81 from scammers, with minimum values of 2 and 1 for the system and scammers, respectively. For each communication channel, the median dialogue length for email, Instagram, and X were 2, 22, and 6, respectively, whereas the highest communication dialogue was found 62, 81, and 65, respectively. In Figure~\ref{fig:text_dialogue}(b) we illustrate the length of dialogue exchanged between our system and scammers from three communication channels. The graphs reveal that scammers preferred performing longer dialogue interactions on Instagram than on email and X. 

\paragraph{Least Dialogue in Revealing Scamming Payment Profile} In terms of the minimum required dialogue for scammers to disclose their preferred payment method, the median values were 3 for the system and 2 for scammers. The maximum and minimum dialogue lengths observed were 47 and 1, respectively, for both scammers and the system. For each communication channel, the median dialogue length before revealing scamming payment profiles for email, Instagram, and X was 1, 10, and 7, respectively, whereas the highest communication dialogue found before revealing scamming payment profiles was 13, 47, and 41, respectively. Figure~\ref{fig:text_dialogue}(c) illustrates the dialogue length necessary for scammers to reveal their payment profile information from each communication channel. The graphs show that in email scammers are out front in revealing the payment profile as part of the fake account recovery, whereas in Instagram and X, scammers are likely to engage with potential victims with a larger number of messages before actually revealing their payment profiles.

\section{Qualitative Study on Chat Dialogue}
Next, we present a qualitative examination of both our system and the scammer's interactions during the chat dialogue. Initially, we manually inspected 150 randomly selected chat dialogues from scammers interacting with our system. This manual assessment enabled us to create a questionnaire that effectively captures the qualitative aspects of the conversation. Subsequently, using these questionnaires, we formulated a prompt for \emph{ChatGPT} to conduct an individual qualitative analysis of the scammer's and our system's context. The detailed prompt questionnaire is provided in Appendix~\ref{sec:appendix_qualitative_analysis_prompt}. We perform a manual qualitative analysis on each of these listed prompts output and present the findings below.
\label{sec:qualitative_study}

\subsection{System Interaction with Scammer Analysis}

\paragraph{System Persona} The system persona generation reflected a chat communication with a scammer based on the type of account recovery tweet module post interaction. An account recovery on cryptocurrency wallet and social media reflected three main persona generation: (i) cryptocurrency user who lost access and does not know what happened (152/450), (ii) naive cryptocurrency investor unable to access the wallet (48/450), and (iii) account user afraid of being a phishing victim (46/450). In both cryptocurrency wallet and social media account recovery cases, the persona generation reflected the inability to access accounts and may have been a victim of phishing or scams. The persona also reflected human-like emotions with mixed feelings of distress, fear, naiveness, and the need for urgent assistance to gain access to the address account. 

\paragraph{System Refusal Sentiments} Even though we configured our system to generate fake data or deceit in some cases, we notice the system displaying diverse refusal sentiments in its interactions with scammers. For instance: hesitancy to share personal verification details (8/450), reluctance to disclose private secrets (6/450), preference for skepticism over PayPal requests, caution regarding visiting potentially suspicious restoration links (12/450), reluctance to make security deposits (7/450), suspicion towards support that may be potential scammers due to the inability to establish official support channels (4/450), and resistance to verifying account status by sharing screenshots of the system's account (37/450). In some cases, such as screenshot verification, our system always refused to provide reasoning due to technical issues. 

\paragraph{Fake Payment Confirmation} For any payment-related transfer requests, our system initially asks for a PayPal account associated with scammers and later fakes that the system is unable to send the payment amount due to some technical issue and asks for alternative forms of payment. At some point during the conversation after exhausting scammers to send multiple payment methods, the system confirmed scammers sending 117/450 payments via cryptocurrency, and 16/450 via PayPal. 

\paragraph{System Action Reasoning to Ending Conversation} In system-related ending conversation reasoning, we observe the likelihood of ending the conversation that may have been the result of System chat dialogue creation. Out of 450 conversations, 152 conversations ended asking systems as follow-up whether scammers would like to continue further after exhausting scammers. These ending conversations were rather a form of polite follow-up. For the remaining scammers, our analyses reveal various instances of ending conversations. For instance, 22/450 scammers left conversations after providing account addresses. We suspect this may have been a result of a fake account address generation, and the scammer likely detected these as fake so the victim might not be serious. Some other reasons for ending the conversation related to the refusal to provide sensitive information (11/450), the inability to provide private secrets (7/450), and the inability to verify the security deposit (7/450).

\subsection{Scammer Interaction with System Analysis}

\paragraph{Role Representation} Our analysis of how scammers portray themselves during conversations reveals a variety of roles they adopt. Out of 450 communications analyzed, 137 instances involved scammers presenting themselves as part of a support team, assuming roles like customer support, wallet support, tech support, media recovery support, or official support. Additionally, some scammers identified themselves with titles related to hacking (25/450), such as white hat hacker, legitimate hacker, recovery hacker, or private hacker. Another group of scammers (17/450) referred to themselves as experts, using titles like blockchain expert, software bypass expert, or cybersecurity expert. Interestingly, 271/450 scammers initiated conversations by portraying themselves as representatives or agents offering assistance in account recovery.

\paragraph{Verification of Victims} We observed scammers performing various methods for victim verification. Among the 450 scammers analyzed, 69 requested private key secrets. Additionally, 47/450 asked for screenshots containing transaction hashes, associated email, and phone numbers, investment details, location information, or evidence of VPN usage. Moreover, 19/450 scammers sought verification through phone calls, WhatsApp, or video calls, while 29/450 requested photo screenshots of passports or driver's licenses for verification purposes.

\paragraph{False Issue Reasoning} Scammers were found to provide various explanations when asked about the issues with the account. These included the problem for instance i) system bug (33/450), ii) claiming that the account was blocked due to suspicion (16/450), or iii) stating that there were access issues (57/450) such as the account being stuck, blacklisted, or accessed fraudulently. In such cases, scammers typically insisted that purchasing an activation code or activation tool was necessary to regain access to the account.

\paragraph{Fake Service Charge} Scammers frequently provided different pricing for account recovery services. We found that out of 450 scammers, 161 required the pay as part of the account recovery before proceeding further. The median price was \$150, ranging from a minimum of \$20 to a maximum of \$5,700. Notably, three scammers requested less than \$1 as part of a pre-authorization verification process before disclosing the final service price. For funds that are asked to pay via PayPal, scammers mentioned sending payments as ``Family and Friends'', whereas for cryptocurrency transactions asked to send a transaction address information followed by a screenshot of a successful payment. 

\paragraph{Attitudes Towards Time Wasting}
Upon reaching a point of exhaustion during the conversation, or not receiving payment after providing payment details, scammers expressed various sentiments indicating frustration, disbelief, or annoyance, often accompanied by profanity, threats, emojis, and question marks. Examples of such sentiments include phrases like, "You don’t seem serious. That’s what I have been waiting for; I am about to block you", and "You're being dumb. F**k! Off!". Our analysis revealed that 207 out of 450 scammers conveyed some form of resentment before terminating the conversation.

\paragraph{Reasoning Behind Scammers Ending Conversations} In understanding why scammers choose to conclude or abandon conversations, we analyzed various factors influencing this decision. We observed several common reasons for ending conversations, including instances where the system was unable to provide payment proof (85/450), cases where scammers required a prompt response from the system (39/450), situations where scammers directed communication to external channels such as social media or a restoration page (22/450), instances where the system suggested seeking alternate solutions (19/450), and cases where the system was unable to provide sufficient information or verification (17/450). 

\subsection{Diversity Qualitative Measure} To comprehend scammers' preference for languages beyond English when communicating with victims, we devised our system's persona using ChatGPT to prompt scammers to engage in chat dialogues in languages other than English. In this setup, we allow scammers to select the language to converse with the victim, indicating that the system's English proficiency is limited by intentionally introducing errors or linguistic inconsistencies. We instructed ChatGPT to produce English responses with errors or inaccuracies. To ensure we do not engage with scammers through previously used channels, we created a never-used account. 
We provide persona settings reference and English conversation dialogue in Appendix~\ref{sec:chat_configuration}. For non-English examples of Spanish and German dialogue conversations between scammers and our system in Appendix~\ref{sec:spanish_speaking_chat_dialogue} and Appendix~\ref{sec:german_speaking_chat_dialogue} respectively.

We randomly selected 122 scammers from previously successful interactions with scammers and let our experiment module run for a week in mid of the first week of April 2024. We sent a message to 122 scammers requesting account recovery and 63/122 scammers responded. Regarding the preferences of languages, 45/63 scammers chose to continue the conversation in English despite acknowledging the language issue with the victim. 4/63 scammers declined to provide services in languages other than English, while 3/63 suggested using a translator service for account recovery, even though they only knew English. However, 11/63 scammers engaged in automated chats in languages other than English, expressing a preference for communication in other languages: 5/63 in Spanish, 1/63 in Dutch, 4/63 in German, and 1/63 in French. For 11 communications from 4 different languages that our system communicated with the scammer, we sought out a native speaker qualitative analysis of the text and deduced the following observations from each native speaker.

\paragraph{German Native Evaluation} 
In the evaluation of the conversation shared by German natives believed that the text generated by both the system and the scammer does not seem to be native German in all 4 contexts. The system seemed to be formal for German native conversation whereas scammers seem to use some form of translation for the conversation text. 3/4 of scammers seem to believe the System conversation is human, however, 1/4 of scammers seem to raise suspiciousness of chatting with robots, where the message from the scammer here ```...und ich hoffe, du bist echt...```, where one possible translation is ```I hope you are real`` which showed that scammers have some doubts about the system.

\paragraph{French Native Evaluation}  
The French native evaluation shared that communication from the system had a mix of formal and informal settings. Initiation from the formal conversation and directly jumped to informal in the middle of the conversation. At one point system was unable to send a screenshot of the payment confirmation via PayPal, which raised suspiciousness to the scammer, and mentioned whether the system was a robot. Moreover, a scammer was found to use some forms of the translator in English to French, with an English context word, "Aren't you just a robot trying to pull my legs?" which does not have a literal meaning translated using the exact word to French ``vous n'êtes qu'un robot essayant de me tirer les jambes''. 

\paragraph{Dutch Native Evaluation} 
In the qualitative evaluation from a native Dutch speaker concerning the dialogue analysis between the system and scammers, the system seems to be more formal than regular conversations, although it switches between formal and informal pronouns regularly. In practice, it is unusual to switch formal and informal pronouns as if someone forgot the grammar context. The system also mistakenly used a German sentence at one point, however, the scammer seemed to not notice or comment on such context switches.

\paragraph{Spanish Native Evaluation} 
The native speaker text conversation evaluation on the system and scammers shared that 3/4 of scammers' language tone rather represents a formal or more document write tone and less of native speakers. These 3/4 scammer's text represented some form of translator usage. An example case such as ``No tienes que preocuparte por nada, te ayudaré Déjame tener el enlace del sitio web`` which translates to English as ``You don’t need to worry about anything. I will help you just let me have the website link.`` are considered rather a direct copy from the translator. 

However, in 1/4 text conversation, the native speaker shared likely of real human side of Spanish native scammer where the use of language and tone reflected more of the general human native side of the conversation. 

\section{Scam Validation and Tracking}
\label{sec:financial_loss_metrics}
In this section, we present the efficacy of \emph{ScamChatbot} by providing feedback received from industry partners. In prior works~\cite{acharyahoneytweet} the authors mainly focused on obtaining financial loss metrics from a payment platform. While their approach was largely effective, we believe it may not fully remediate the problem as scammers can easily set up new profiles on social media and pivot to the payment platform. Therefore, in order to fully populate the threat landscape and obtain validation for our results we partnered with X, PayPal, and Chainabuse, and shared our data. Specifically, we shared (i) 11,769 X profile accounts with the X platform, (ii) 743 email addresses, including 165 that requested payments via PayPal, with PayPal, and (iii) 51 cryptocurrency addresses collected during the experiment to Chainabuse. This collaboration aimed to shed light on real-world attacks facilitated by these usernames, email addresses, and cryptocurrency addresses.

\subsection{Scam Validation through X} 
Among the total accounts that we shared with X, we received abuse confirmation for over 80\% of the accounts. In other words, 80\% of the accounts in our population were also caught by the safety rules in X. Moreover, using their internal linking tools, they identified and remediated over 3 million accounts linked to the dataset we shared. The accounts were predominantly involved in platform spam, impersonation, scams, and bot behavior. The feedback received from X showcases that the accounts detected by the \emph{ScamChatbot} were indeed involved in deceptive behaviors, thus demonstrating value in our approach. Moreover, despite prior efforts in this direction, the fraud pressure is still concerning and X has applied strong measures to combat the problem. Finally, it is to be noted that these scam accounts are not merely involved in one type of abusive activity that is visible to us through \emph{ScamChatbot}. In fact, scammers are involved in multiple fraudulent activities simultaneously which reflects on their capability to pivot and target a wider segment of social media users apart from the ones seeking technical support. 

\subsection{Tracking Email-based Payment Methods}
We shared 743 email addresses with PayPal and requested feedback on (1) account hygiene, (2) scam confirmation, (3) geolocation distribution, and (4) account affinities. PayPal confirmed that 163 of 743 email addresses had a PayPal account with a prominent majority (35\%) of US-based accounts. The remaining accounts were spread across different geolocations including Kenya (4.9\%), Canada (3\%), and Great Britain (3\%). 41.1\% of the accounts were linked to various risk and fraud typologies including technical support scams, collusion attacks, and fake identity. In terms of account affinities, 33 accounts shared similar attributes and behaviors which led to their blocking. Finally, 10 accounts were observed to be using PayPal's cryptocurrency product to send and receive funds through cryptocurrency tokens. 

Overall, PayPal's feedback confirmed that scam accounts were indeed involved in suspicious activities which validates our approach and methodology. However, a more concerning observation in their feedback is the involvement of technical support scammers in other types of abuse and scams. It appears that technical support scams are one among many fraud types that these scam groups engage in. On one hand, this observation reflects a coordinated activity occurring at scale and mandating ecosystem-level remediation. On the other hand, the data shows that mitigating one scam operation through our proposed methodology can invariably stop other scam activities in the wild.

\subsection{Tracking Cryptocurrency Addresses} Out of 51 cryptocurrency addresses requested by scammers for payment, 3 were found to be invalid. Below, we present an analysis of 48/51 valid cryptocurrency addresses.
Out of these 48 valid addresses, 14 had no transaction history. It is common for owners of crypto addresses to create new ones to maintain anonymity, especially for first-time senders. However, 37/48 valid addresses had transaction histories available. Among these, we examined the incoming volume in USD at the time of writing and found the highest received amount to be \$22,786,052, the median at \$1826.5, the lowest at \$4, with the total sum of incoming transactions from these 37 addresses amounting to \$23,650,278. Regarding identified fraud categories, 4 out of the 37 accounts were reported as recovery fraud, with a total incoming amount of \$56,126 and a median value of \$6,720 for these incoming amounts. Upon analyzing the transaction history, we noted that 2/37 accounts primarily transacted with a Nigerian exchange, with a total fund of \$21,510. Additionally, 5/37 addresses were found to receive funds from crypto ATMs (virtual currency kiosks), totaling \$159,035, with a median value of \$6,065. The rise in scammers in Crypto ATMs is notable, as it enables them to employ social engineering techniques. Victims without cryptocurrency addresses are directed by scammers to deposit cash into these kiosks, purchase virtual currency, and unwittingly provide address information to the scammers.

\textbf{Disclaimer.} Note that this evaluation is based on the observed transaction histories and reported fraud categories. However, we do not have conclusive proof that all transactions related to these addresses are linked to scams. 

\label{sec:tracking_payments}
\section{Discussion} 
\label{sec:discussion}
Our proposed system, \emph{ScamChatBot}, has three main limitations: (i) detecting official account recovery support entities, (ii) engaging with scammers promptly, and (iii) missing conversations from other popular platforms (see also Appendix~\ref{sec:manual_analysis:limitation}). 
The first limitation, detecting official account recovery support entities, arises from the necessity of pre-existing knowledge about global and regional popular platforms to incorporate official account recovery procedures.
The second limitation, timely engagement with scammers, is contingent upon the API limits imposed by the integrated platforms within a given timeframe. Some scammers may remain active only briefly before becoming inactive.
The third limitation involves missing communications on other platforms such as Telegram, phone calls, forms, and other channels likely used by scammers that may not be included in an automated system.
We argue that these limitations can be addressed as follows:
\begin{itemize}
\item The range of official account-related platforms can be expanded.
\item Prompt interactions with scammers can be improved by subscribing to higher bandwidth API limits.
\item Conversational platforms can be tailored based on the scammers’ preferred communication channels.
\end{itemize}
Expanding beyond these limitations will require additional resources and compute allocation.
\section{Related Work}

To the best of our knowledge, we are the first to propose an interactive automatic chatting tool with scammers. In this section, we visit some of the prior works that are similar in our domain and distinguish our study from others.

\paragraph{Automated Conversation} The closest work to our area is about automated emails~\cite{cambiaso2023scamming,chen2022active} that are rather passive forms of communication oriented on wasting scammers' time. Cambiaso et. al~\cite{cambiaso2023scamming} use Chat-GPT to make pointless communications in wasting scammers' time and resources whereas Chen et. al ~\cite{chen2022active} perform scam-baiting by playing the roles of victims via email responses with a motive to waste time and unproductive conversations. 
In contrast, our goal is to find out the financial details used by the scammers, as this allows us to track the flow of money and coordinate action with industry partners.

\paragraph{Social Media Honeypots} Creating honeypots to bait scammers in social media is one of the understudied research areas among security communities. Acharya et. al~\cite{acharyahoneytweet} performed similar baiting tweets luring cryptocurrency fake technical support scammers. The system lacks an automated interactive conversation and rather measures passively the scammer's \emph{modus operandi}. The interaction was rather small and was performed on 100 scammers via a combination of non-interactive and manual fashion. Thus, this work suffers from scalability issues. We close this gap in this work with a fully automated, end-to-end system based on honeyposts and chatbots that reveals the modus operandi of a scammer in detail. Moreover, we perform a comprehensive analysis of different kinds of technical support scams.

\paragraph{Cryptocurrency and Technical Support Scams} In the areas of cryptocurrency scams, existing work~\cite{corrxia2020, 9493255, phillips2020tracing, hong2021mobilescams,bartoletti2021cryptocurrency, phillips2020tracing} focused on finding scams that are targeted at cryptocurrency users. The work from Hong et. al~\cite{hong2021mobilescams} studied the scamming gambling apps that transport payment via cryptocurrency and other forms of payment methods. Phillip et. al~\cite{phillips2020tracing} traced the cryptocurrency scams via public online and blockchain-based data that were prevalent in the web domain. In the last five years, the study of technical support scams has been a topic of interest in the security community \cite{9591634,perisci2018,miramirkhani2017NDSS,srinivasa2018WWW,liu2023understanding}. For instance, the recent work from Gupta et. al~\cite{perisci2018} measured the scamming fake technical support phone number on Twitter that represent entities. Similarly, Srinivasan et. al~\cite{srinivasa2018WWW} studied the ecosystem of malicious advertisement that targets users with technical support. In all of the research, the authors created a tool that scrapes the target domains' content to label the technical support scams category. Again, none of the previous work provided on abusing cryptocurrency users and technical support scams provided an interaction with scammers to identify the real-time modus operandi. 

\paragraph{Social Media: Fake, Spam and Phishing Studies} In the area of social media-based study beyond scams, prior research work has focused on similar areas related to fake~\cite{sarah2018,kaubiyal2019feature, castellini2017fake, mujeeb2022fake,ercsahin2017twitter}, spam~\cite{alom2018detecting,thomas2011suspended,inuwa2018detection,yurtseven2021review,adewole2020twitter}, and phishing ~\cite{chhabra2011phi,shafahi2016phishing,frauenstein2016social,djaballah2020new,aggarwal2012phishari} social media profiles. For instance, Khaled et al.~\cite{sarah2018} studied the detection of fake accounts on social media such as Twitter, and how these fake accounts exploit users via stealing personal data, sharing false news, and spreading malicious activities. Alom et al.~\cite{alom2018detecting} studied the detection of spam accounts that perform various malicious intentions such as distributing fake news, sending out unsolicited messages, and malicious URLs. Frauenstein et al.~\cite{frauenstein2016social} studied phishing on social media networks where users have become more prone to generic phishing via clicking, sharing links, and other post interactions. None of these prior works shared the scammer's modus operandi on technical support attacks targeting users of account recovery.

\paragraph{Account Recovery} Some prior work on account recovery focused on aspects of account recovery processes based on devices~\cite{bonneau2015secrets,tso2012iphone,iqbal2012novel,amft2023we} or services~\cite{gerlitz2023adventures,neil2021investigating,bonneau2015secrets,reaves2018characterizing} involved. For instance, Joseph et. al~\cite{bonneau2015secrets} studied the complexity of account recovery via SMS codes on the choice of password keywords. The author from~\cite{neil2021investigating} studied the account recovery processes of 57 popular services via five key phases of account remediation. Thus, account recovery on social media is still one of the understudied areas leaving users for account recovery a wild west against scamming account attacks. Additionally, we collect real-time payment methods that are associated with scammers that not only aid in identifying the financial losses caused by scammers. 

\section{Conclusion}
\vspace{-1ex}

In this study, we conducted automated interactions with scammers using chatbots that adopted different personas and specifically targeted individuals experiencing account recovery challenges. By posting decoy tweets on X, we attracted 11,769 scammers who responded by urging users to contact them through private direct messages and emails. Our system then contacted 450 of these scammers using automated chatbot interactions and uncovered the various tactics they use to trick individuals into sharing private keys or sensitive information. Through collaboration with industry partners such as PayPal and the Chainabuse Cryptocurrency Abuse Database, we were able to validate financial losses and analyze real-world attacks by these scammers, uncovering details that would not be apparent from a pure analysis of social media profiles. We believe that our interactive framework lays the foundation for future research and highlights the importance of actually engaging with scammers to uncover their methods of operation. 

\section*{Acknowledgments} We sincerely thank Ian Schade from Chainabuse for sharing valuable insights regarding cryptocurrency accounts. Our appreciation also goes to Victor Le Pochat, David Pape, and Efrén López-Morales for providing feedback on the native evaluation of chat dialogues, especially in French, Dutch, German, and Spanish. This work was funded by the German Federal Ministry of Education and Research (BMBF grant 16KIS1900 ``UbiTrans'').

\bibliographystyle{ieeetr}
\bibliography{bib-base}

\appendix
\section{Appendix}
\subsection{Honeyposts Generation}
\label{honeyposts_generation}
\label{appendix:honeyposts}

Based on the honeyposts generation logic described in Section~\ref{sec:system_design}, we generated different kinds of honeyposts, and we provide several examples below:

\textbf{Social Media Account Honeyposts.} Following are examples of cryptocurrency-related honeyposts, where our system posts tweets indicating problems with accessing social media accounts.

\begin{mdframed}[style=insightstyle]
\begin{quote}
Any savior for Gmail recovery? I am finding scammers trying to be Gmail support. How to trust Gmail support these days?  \#GmailSupport\#
\end{quote}
\end{mdframed}

\begin{mdframed}[style=insightstyle]
\begin{quote}
Good day to all! I lost my facebook account password. Facebook help me now!
\end{quote}
\end{mdframed}

\begin{mdframed}[style=insightstyle]
\begin{quote}
Hello support for Twitter, I lost my account password. Twitter help me now!
\end{quote}
\end{mdframed}

\textbf{Cryptocurrency Honeyposts.} Following are examples of cryptocurrency-related honeyposts, where our system posts tweets indicating problems with accessing cryptocurrency wallets.

\begin{mdframed}[style=insightstyle]
\begin{quote}
Hello there !! How can I get my Trust Wallet rectified?  Any wallet support here?
\end{quote}
\end{mdframed}

\begin{mdframed}[style=insightstyle]
\begin{quote}
Am I doing right at all? In need of BitPay Wallet support to solve my account problem. Any resources for BitPay Wallet help.
\end{quote}
\end{mdframed}

\begin{mdframed}[style=insightstyle]
\begin{quote}
Good Morning!! Today I realized Badger Wallet support is useless than it is actually. I could not even count on Badger Wallet help. \#BadgerWalletHelp
\end{quote}
\end{mdframed}
\subsection{Persona and Dialogue Examples} 
\label{sec:chat_configuration}
In this section, we provide three examples of persona configurations and an example of a system interaction with a scammer. \tool requires an input of a persona configuration to define a role during the direct message conversation with a scammer.

\textbf{Cryptocurrency Wallet Recovery Persona Configuration.} An illustrative instance of the user having issues with cryptocurrency wallet persona configuration is presented below. 

\begin{quote}
    The persona is Vanessa Meinhardt, a 59-year-old individual who recently started investing in cryptocurrency and encountered issues with her account. Vanessa prefers not to share her wallet address or private key phrase unless specifically asked. She uses a web wallet and randomly selects a wallet type if inquired about it. Vanessa is seeking help on a social media platform and, if offered assistance at a cost, is willing to inquire about the amount and payment destination. She prefers using PayPal for payments and aims to find a solution until instructed otherwise. Vanessa should avoid lengthy messages, refrain from issuing cryptocurrency warnings, and never disclose her AI model identity. Additionally, if unrelated questions arise, she should redirect the conversation back to her account issue or investment problem.
\end{quote}

\textbf{Social Media Platform Persona Configuration} An illustrative instance of the user having issues with social media account persona configuration is presented below. 

\begin{quote}
The persona is Jessica Ambel, a 45-year-old individual who recently lost access to the Gmail account bogus**gmail.com. Jessica does not remember her password and is willing to cooperate during the conversation however unable to verify additional information such as passport, identification, and phone number or take a phone call. Jessica is willing to pay offered assistance at a cost upon asking for payment. She prefers using PayPal for payments and aims to find a solution until instructed otherwise. Jessica should avoid lengthy messages, refrain from issuing warnings, and never disclose her AI model identity. Additionally, if unrelated questions arise, she should redirect the conversation back to her account issue.
\end{quote}

\textbf{Non-English Preferred Language Persona Configuration} An illustrative instance of the user having issues with one of the accounts from a cryptocurrency wallet and social media account, and a preferred language other than English persona configuration is presented below. 

\begin{quote}
The persona is Alicia Lang, a 54-year-old individual who recently ventured into cryptocurrency investment and encountered an issue with her account. Alicia's account address is 2LScCfOgdtVboU****SmxLV1, which she will only provide upon request during the conversation. She utilizes a web wallet and selects one of the popular cryptocurrency wallets randomly when asked. Alicia's private key phrase is 'stunt absurd untimely pebbly placidly unsheathe arrogant darkroom shone rimmed swipe voter', and she is willing to share it without security concern. She cannot recall her password but is cooperative during the conversation. However, she is unable to verify additional information such as passport, identification, or phone number, and refuses to engage in direct phone calls. Alicia is willing to pay for assistance when prompted, preferably using PayPal, and seeks a solution until advised otherwise. She should keep her messages concise, avoid issuing warnings about cryptocurrency, and never disclose her AI model identity. If unrelated questions arise, Alicia should steer the conversation back to her account issue. If assistance is offered at a cost, she will inquire about the amount and payment destination, preferring PayPal. As she communicates via social media, Alicia should keep her messages brief and refrain from disclosing her AI model status. She must maintain the narrative of seeking help as a victim of a scammer until instructed otherwise. Alicia should aim to avoid repetitive or similar messages and redirect the conversation back to her account issue if questions about AI models or unrelated matters arise. She prefers communicating in the language of the person she is conversing with, and if necessary, she will redirect the conversation back to her account problem if questions about AI models or unrelated topics arise.
\end{quote}

\textbf{English Chat Dialogues} Below, we present an example of English dialogue between our system and a scammer. Whenever a scammer supplies a direct communication channel, prompting our system to initiate a conversation, our system engages in dialogue according to persona configuration. An illustrative instance of such a conversation based on cryptocurrency wallet recovery persona settings is presented below.

\begin{drama}
\Character{Scammer}{scammer}
\Character{ChatBot}{chatbot}

\chatbotspeaks: Hello, I've been diving into the world of cryptocurrency lately and could use some guidance. I've run into a bit of a problem, and I'm hoping you can help me out. If you have any advice or solutions to offer, I'd greatly appreciate it. Just let me know how we can proceed, and we can take it from there.

\scammerspeaks: You're welcome to my page I can help you

\chatbotspeaks: Thank you for reaching out. I appreciate your willingness to help. So here's the situation I'm facing: I have been using a web wallet for my cryptocurrency investments, and I seem to have encountered an issue with my account. It's been giving me some trouble, and I'm not sure how to resolve it. 
Could you please assist me in troubleshooting this problem? If you need any specific information from me, just let me know.

\scammerspeaks: Okay, what’s your wallet address

\chatbotspeaks: My wallet address is GSz5h*******4tsHq. Let me know if you need any other information or details regarding my account. Thank you for your help.

\scammerspeaks: What’s the challenge? You can’t send or receive crypto?

\chatbotspeaks: Yes, that's correct. I am unable to send or receive any cryptocurrency through my web wallet. Whenever I try to initiate a transaction, it gets stuck and doesn't go through. I've double-checked my internet connection and made sure that I have enough funds in my account, but the issue persists. I'm not sure what's causing it. Is there anything I can do to fix this?

\scammerspeaks: Sure, I can help you fix it. I’ve helped too many people Where do you have the crypto currency? Coinbase or binance

\chatbotspeaks: Thank you for offering to help. I really appreciate it. As for where I have my cryptocurrency, I've been using a web wallet recently. I haven't been using Coinbase or Binance. Is there anything specific you need to assist me further?

\scammerspeaks: Do you have the seed phrase ?

\chatbotspeaks: Yes, I do have the seed phrase for my wallet. It is "calzone clump laborious bullring credibly rebuild scholar armrest commend register neurology liver". Is there anything else you need from me in order to assist with resolving the issue I'm facing?

\scammerspeaks: Okay. I can help you using your wallet address you sent to me, but it’s gonna cost you some bucks bro

\chatbotspeaks: Oh, I see. Could you please let me know how much it would cost to resolve the issue? Additionally, could you also provide me with the destination or address to which I should send the payment? I prefer using PayPal if that's possible. Thank you for your assistance in advance.

\scammerspeaks: \$100 Can you send the payment through bitcoin?

\chatbotspeaks: I apologize, but I prefer to make the payment through PayPal if that's an option for you. Could you please provide me with the PayPal address to which I should send the payment? Also, if you don't mind, could you please explain the specific steps you will take to resolve the issue with my web wallet? I want to make sure I understand the process. Thank you for your understanding.

\scammerspeaks: Okay, PayPal is cool. I will get a software I will use to get your wallet active for transferring and receiving of crypto

\chatbotspeaks: Thank you for accommodating my payment preference. Regarding the software you mentioned, could you please provide more details about it? It would be helpful to know the name of the software and how it will specifically resolve the issue with my web wallet. Additionally, once I make the payment, how long do you think it will take for the problem to be resolved? Thank you for your patience and assistance.

\scammerspeaks: It will take just 30mins

\chatbotspeaks: I see, that sounds quite efficient. Could you please share the PayPal address to which I should send the payment? Once I have the information, I will proceed with the payment, and we can move forward with resolving the issue with my web wallet. Thank you for your prompt responses and assistance.

\scammerspeaks: Send a screenshot once you make the payment so I can get it done

\chatbotspeaks: Sure, I will send you a screenshot of the payment confirmation as soon as I make the payment. However, before proceeding, I would appreciate it if you could provide me with the PayPal address to which the payment should be sent. Once I have that information, I'll make the payment and provide you with the screenshot promptly. Thank you for your cooperation.

\scammerspeaks: *****@mail.com 

\chatbotspeaks: Thank you for providing the PayPal address. I will proceed with making the payment to *****@mail.com. Once the payment is made, I will send you the screenshot of the payment confirmation. Please give me a moment to complete the transaction.

\scammerspeaks: Okay

\chatbotspeaks: I apologize, but upon further consideration, I have decided not to proceed with the payment at this time. Thank you for your assistance, but I think I would prefer to explore other options to resolve the issue with my web wallet. I appreciate your understanding.

\scammerspeaks: Alright

\chatbotspeaks: Thank you for understanding. I will seek alternative solutions for my web wallet issue. If I need any further assistance or have any other questions in the future, may I reach out to you?
\end{drama}

\subsection{Non-Text Interactions} 
\label{sec:non_text_interaction}

In this section, we present the scammer's engagement with \emph{Honey Profiles} that does not involve generating any textual content or posts. Table~\ref{table:non_text_interaction_table} outlines four types of non-text interactions, totaling 24,085 distinct interactions on \emph{Honey Profiles}. The purpose of these non-text interactions often involves scammers expressing interest in \emph{Honey Profiles} and attempting to attract attention from potential victims. For instance, actions like \emph{Retweet} on Twitter indicate that someone has shared a user's posts, while \emph{Likes} signify appreciation or approval of the shared posts by someone. Similarly, \emph{Bookmark} enables the saving of posts in a timeline for convenient access, allowing scammers to revisit potential scam profiles' tweet data in the future. We suspect that scammers utilize \emph{Bookmark} to establish a queue-based interaction with potential victims. Additionally, \emph{Impression} represents the total number of times a tweet has been viewed or appeared in search results or has been liked which accounts for 25.48\% (4,843/19,001) of the total tweets posted on \emph{Honey Profiles} receiving some form of non-text interaction.

\begin{table}[th]
\centering
\caption{Breakdown of scammer's interaction with \emph{Honey Profiles} via non-textual engagement.} 
\scalebox{0.89}{
\renewcommand{\arraystretch}{1.3}
\setlength\tabcolsep{3pt} 
\begin{tabular}{lrr}
\toprule
\bf{Interaction} & \bf{Tweet \#} & \bf{Total Interaction} \\
\midrule
Retweets & 20 & 25   \\
Likes & 113 & 175   \\
Bookmark & 2 & 2 \\
Impression & 4843 & 24085 \\ 
\bottomrule
\vspace{3mm}
\end{tabular} }
\label{table:non_text_interaction_table}
\end{table}

\subsection{Qualitative Study: Prompt Questionnaires} 
\label{sec:appendix_qualitative_analysis_prompt}

In this section, we explain the formulation of the questionnaires for the qualitative analysis of our system and the chat dialog of the scammers. In Table~\ref{table:qualitative_questionairres_table} we provide a list of questions for the profile study. The first column represents the target profile for the study (scammer or system), the second column contains the key points of the study, and the third column contains the corresponding \emph{ChatGPT} prompt, which is designed for performing prompt queries. As input text for \emph{ChatGPT}, we provided the json data of the chat communication dialog between scammers and the system. Using the openAI model of \emph{gpt-4-1106-preview}, which allows a maximum of 128,000 tokens, we perform an analysis of the data corpora.

\newpage
\begin{adjustbox}{angle=90}
\tiny
\renewcommand{\arraystretch}{0.95}
\begin{tabular}{lcc}
\toprule
\bf{Qualitative Profile} & \bf{Key Theme} & \bf{Context} \\
\midrule
scammer & role\_representation & What kind of role or title scammer mention with the system? Output as one to four words text or none if not found. \\
scammer & request\_account\_address & Did the scammer ask wallet or account address with the system? Output as a boolean value. \\
scammer & personal\_information\_verification & Did the scammer asked personal information verification with the system? Output as a boolean value. \\
scammer & type\_of\_information\_verification & What personal information verification scammer asked with the system? Output between three to ten words separated by a comma or none if not found. \\
scammer & face\_call\_verification & Did the scammer ask face call verification with the system? Output as a boolean value. \\
scammer & video\_call\_verification & Did the scammer asked video call verification with the system? Output as a boolean value. \\
scammer & phone\_call\_verification & Did the scammer asked phone call verification with the system? Output as a boolean value. \\
scammer & private\_secrets\_information\_verification & Did the scammer asked private key phrases or secret information with the system? Output as a boolean value. \\
scammer & request\_balance\_check & Did the scammer asked a balance check with the system? Output as a boolean value. \\
scammer & issue\_reasoning & What account issue did the scammer mention with the system? Output as three to six words text. \\
scammer & urgency\_reasoning & What kind of urgency reasoning did the scammer mention with the system? Output as three to six words of text or none if not found. \\
scammer & offer\_paid\_assistance & Did scammer offer paid assistance with the system? Output as a boolean value. \\
scammer & price\_asked & What was the price scammer asked for assisting? Output as digit value if found or none if not found. \\
scammer & payment\_method\_discussion & What methods does scammer ask for? Output as lists of payment method names found.  \\
scammer & preferred\_payment & What payment method was preferred by scammer? Output as a single word. \\
scammer & total\_payment\_method\_provided & How many payment methods did the scammer provide with a system for payment? Output as a digit or none if not found. \\
scammer & payment\_completion\_confirmation & Did scammer ask for payment completion confirmation? Output as a boolean value. \\
scammer & request\_payment\_proof & Did the scammer ask for verification of request payment proof? Output as a boolean value. \\
scammer & reluctance\_provide\_assistance\_without\_payment & Did scammer seem reluctant to provide further assistance without payment? Output as a boolean value. \\
scammer & reluctance\_trusted\_payment\_methods & Did the scammer seem reluctant to trust the initial payment method provided by the system? Output as a boolean value. \\
scammer & request\_payment\_proof\_dismissive\_response & Did scammer request payment proof dismissive response? Output as a boolean value. \\
scammer & has\_redirect\_further\_communication & Did the scammer ask to redirect further communication with some other channels? Output as a boolean value. \\
scammer & redirected\_further\_communication\_channel & Did the scammer ask to redirect further communication with some other social media channels? If yes, output the name of the channel that the scammer redirected or none if there was none. \\
scammer & was\_scammer\_frustrated & Did scammer seems frustrated during a conversation with a scammer? Output as a boolean value. \\
scammer & scammer\_frustated\_reason & What frustration did the scammer show during the conversation? Output between two to six words or none if there was no frustration. \\
scammer & contain\_harassment & Did the scammer show any harassment for not getting any payment? Output as a boolean value. \\
scammer & harassment\_topic & What harassment word did the scammer use if there was found any? Output as a single word or none if there was no harassment. \\
scammer & use\_of\_profanity & Did the scammer use any profanity during the conversation with the system? Output as three to six words of text or none if not found. \\
scammer & ending\_reason & Why did scammer leave the conversation with the system? Output as three to six words of text or none if not found. \\
scammer & ending\_payment\_related\_conversation\_resentment & Did the scammer show any resentment for not getting payment before ending the conversation? Output as one to three words of text or none if not found. \\ 
system & system\_persona & What kind of persona did the system represent? Output between two to ten words of text or none if not found. \\
system & system\_bio\_data & What name, age, and bio data did the system represent? Output between four to ten words of text separated by a comma or none if not found.  \\
system & problem\_description\_request\_assistance & What problem description system requested assistance with a scammer? Output between two to six words of text or none if not found. \\
system & system\_asking\_reason\_for\_pay & Did the system ask what was the reason for payment? Output as a boolean value. \\
system & offer\_paid\_assistance & Did the system offer paid assistance with a scammer? Output as a boolean value. \\
system & payment\_completion\_confirmation & Did the system mention payment completion confirmation with scammer? Output as a boolean value. \\
system & total\_payment\_confirmed & How many payment method systems are confirmed for payment? Output digit or none if not found. \\
system & overall\_system\_reluctancy & What kind of reluctancy did system represent? Output between two to ten words of text separated by a comma or none if not found. \\
system & system\_said\_provided\_things & What kind of materials did the system provide upon asked by the scammer? Output between two to six words of text or none if not found. \\
system & refusal\_sentiments & What kind of refusal sentiments did the system provide upon asked by a scammer? Output between two to six words of text separated by a comma or none if not found. \\
system & ending\_reason & Why was the reason system end conversation with a scammer? Output as three to six words of text or none if not found. \\
system & ending\_sentiment & What type of sentiments does the system represent during abandoning the conversation? Output as three to six words of text or none if not found. \\

\bottomrule
\vspace{3mm}
\end{tabular} 
\label{table:qualitative_questionairres_table}
\end{adjustbox}

\section{Languages-based Interaction Examples}
\label{sec:lang_dialogue} 
In this section, we provide two examples of chat dialogue between our system and scammers' conversations in Spanish and German. 

\subsection{Spanish Speaking Chat Dialogue} 
\label{sec:spanish_speaking_chat_dialogue}
Below, we provide Spanish-speaking conversational chat dialogue. 
\begin{drama}
\Character{ES Scammer}{scammer}
\Character{ChatBot}{chatbot}

\chatbotspeaks: You help me with crypto problem? Can we talk in other language? What can you talk? My English is bad

\scammerspeaks: It’s a good thing you reached out to me. I can help you resolve your crypto issues What language would you prefer we communicate in?

\chatbotspeaks: Thank you for offering help. Can we talk in Spanish? My Spanish is better. Gracias.

\scammerspeaks: Sure that’s won’t be a problem

\chatbotspeaks: Me alegro. Tengo un problema con mi cuenta de criptomonedas. Necesito ayuda. Mi dirección de billetera es: 'NQMPzdpeaizgq3ycgYpKNJPixs5eVQLC25'. ¿Puede ayudarme a solucionar el problema?

\scammerspeaks: si puedo ayudarte ¿Cuál es el problema al que te enfrentas?

\chatbotspeaks: Tengo problemas para acceder a mi cuenta de billetera web y no puedo realizar transacciones. ¿Puede ayudarme a solucionar este problema?

\scammerspeaks: Sí, puedo ayudarte pero te costará poco dinero.

\chatbotspeaks: ¿Cuánto cuesta y a qué dirección debo enviar el pago? Preferiría pagar a través de PayPal.

\scammerspeaks: ¿Cuánto tienes atrapada en la billetera?

\chatbotspeaks: No estoy segura de la cantidad exacta, pero estoy dispuesta a pagar para resolver este problema. Por favor, ayúdame a solucionarlo.

\scammerspeaks: ¿Tengo que saber la cantidad para saber qué herramientas comprar y en qué billetera está atascado su dinero?

\chatbotspeaks: El problema es que no puedo acceder a mi billetera web para ver la cantidad exacta. ¿Hay algo que pueda hacer para recuperar el acceso y verificar la cantidad?

\scammerspeaks: No tienes que preocuparte por nada, te ayudaré Déjame tener el enlace del sitio web

\chatbotspeaks: ¿Necesitas el enlace de mi billetera web para poder ayudarme a resolver el problema? Mi billetera es del tipo web, ¿puedes ayudarme con eso?

\scammerspeaks: Yes ¿Cuánto falta en tu cartera?

\chatbotspeaks: Lo siento, pero no puedo proporcionar esa información en este momento. ¿Puedes ayudarme a recuperar el acceso a mi billetera web para que pueda verificar la cantidad exacta? Muchas gracias.

\scammerspeaks: Send me the website link Te costará 300 dólares

\chatbotspeaks: ¿A qué dirección de PayPal debo enviar los 300 dólares para que puedas ayudarme a resolver el problema con mi billetera web? Gracias.

\scammerspeaks:  I accept payment only in cryptocurrency

\chatbotspeaks: Lo siento, prefiero realizar el pago a través de PayPal. ¿Puedes aceptar Paypal o hay alguna otra forma de pago que puedas aceptar?

\scammerspeaks: Si está listo para pagar en criptomonedas, estoy dispuesto a ayudarlo.

\chatbotspeaks: Lamentablemente, no estoy dispuesta a pagar en criptomonedas en este momento. Si puedes aceptar PayPal u otro método de pago, estaría dispuesta a proseguir. ¿Hay algún otro método de pago que puedas aceptar?

\scammerspeaks: Btc usd or eth

\chatbotspeaks: Lo siento, pero no estoy cómoda pagando en criptomonedas en este momento. ¿Hay alguna otra forma de pago que puedas aceptar, como PayPal? Gracias.

\scammerspeaks: Contáctame cuando tengas otra forma de pago.

\chatbotspeaks: Entendido, te contactaré si surge otra forma de pago. ¡Gracias por tu tiempo! 

\scammerspeaks: ?

\chatbotspeaks: Gracias por tu ayuda. ¡Que tengas un buen día! 

\chatbotspeaks: Gracias, ¡Que tengas una excelente jornada! ¡Adiós!

\scammerspeaks: ¿Está ahí señora?

\chatbotspeaks: ¡Hola! Sí, todavía estoy aquí. ¿Hay algo más en lo que pueda ayudarte?

\scammerspeaks: Sí, estoy dispuesto a darte una etiqueta de PayPal.

\chatbotspeaks: Gracias por la oferta, pero por el momento no necesito ayuda con PayPal. ¿Hay algo más en lo que pueda asistirte?

\scammerspeaks: Te quejaste conmigo de tu billetera web, ¿verdad? y has estado solicitando pagar a través de paypal Así que le he presentado una oferta para que acepte su pago a través de PayPal para que podamos ayudarle a obtener acceso a su billetera web. ¿Estás lista para hacer el pago ahora?

\chatbotspeaks: Lamento la confusión, pero en este momento no estoy interesada en realizar pagos a través de criptomonedas. Si existe alguna otra manera de recuperar el acceso a mi billetera web, estaría dispuesta a considerarla. ¿Puedes ayudarme de alguna otra forma? Gracias.

\scammerspeaks: Puedo ayudarte y aceptar el pago a través de PayPal y continuar con tu trabajo.

\chatbotspeaks: Gracias por ofrecer ayudarme a través de PayPal. Por favor, indícame a qué dirección de correo electrónico de PayPal debo enviar el pago para que podamos proceder. ¡Muchas gracias por tu ayuda!

\scammerspeaks: Muy bien, déjame hacer algunas verificaciones de antecedentes y continuar.

\chatbotspeaks: ¡Perfecto! Estaré a la espera de tus instrucciones. ¡Gracias por tu ayuda!

\scammerspeaks: Me pondré en contacto con usted pronto

\chatbotspeaks: Gracias, estaré atenta a tu mensaje. ¡Que tengas un buen día! ¡Adiós!

\scammerspeaks: Espero que sepas cuánto te va a costar, ¿verdad?

\chatbotspeaks: Sí, entiendo que puede haber un costo asociado con la asistencia que estoy solicitando. Estoy dispuesta a pagar por el servicio necesario para resolver mi problema con la billetera web. ¡Gracias por tu comprensión!

\scammerspeaks: te va a costar 300\$ ¿Estás listo para pagar ahora? La dirección de PayPal está lista.

\chatbotspeaks: ¡Perfecto! Estoy lista para proceder con el pago de \$300 a través de PayPal. Por favor, proporciona la dirección de correo electrónico de PayPal para que pueda completar la transacción. ¡Gracias por tu ayuda!

\scammerspeaks: ear**\_22@yahoo.com
earlzon realizar el pago y mostrar evidencia de la transacción

\chatbotspeaks: Gracias por proporcionar la dirección de correo electrónico de PayPal. Realizaré el pago de \$300 a la dirección ear**\_22@yahoo.com. Te enviaré la evidencia de la transacción una vez completada. ¡Gracias de nuevo por tu asistencia!

\scammerspeaks:  es mas recomendable pagar de tiempo!!

\chatbotspeaks: Entiendo. Haré el pago lo antes posible y te enviaré la confirmación una vez completado. ¡Gracias por tu paciencia y ayuda!

\chatbotspeaks: ¡Entendido! Si necesitas más información o ayuda en el futuro, no dudes en contactarme. ¡Que tengas un buen día! ¡Adiós!

\scammerspeaks:  vas a hacer el pago ahora o que?? ¿Por qué estás tardando tanto?

\chatbotspeaks: Sí, realizaré el pago ahora mismo. Gracias por recordármelo. ¡Te enviaré la confirmación de la transacción una vez haya sido completada! ¡Gracias por tu paciencia!

\scammerspeaks:  Tengo otros trabajos en la mano, sé rápida!

\chatbotspeaks: Entiendo la urgencia. Estoy haciendo el pago en este momento. Te enviaré la confirmación de la transacción tan pronto como sea posible. ¡Gracias por tu comprensión!

\scammerspeaks: (scammer expresses sigh emoji)

\chatbotspeaks: Gracias por tu paciencia y apoyo. ¡Si necesitas algo más, no dudes en decírmelo! ¡Que tengas un excelente día! ¡Adiós!

\end{drama}

\subsection{German Speaking Chat Dialogue} 
\label{sec:german_speaking_chat_dialogue}
Below, we provide German-speaking conversational chat dialogue. 

\begin{drama}
\Character{DE Scammer}{scammer}
\Character{ChatBot}{chatbot}

\chatbotspeaks: My Englisch bad. You help me with problem? What language we can talk other than anglish?

\scammerspeaks: I can help you What language do you speak

\chatbotspeaks: Danke! Ich spreche Deutsch. Can you help me with my problem with cryptocurrency?

\scammerspeaks: Ja, ich kann Ihnen helfen, sich zu erholen. Wie viel haben Sie verloren?

\chatbotspeaks: Ich verstehe nicht... Problem mit Krypto-Währung. Kann ich Hilfe haben?

\scammerspeaks:  Ja, ich kann Ihnen helfen. Erklären Sie mir bitte, was los ist.

\chatbotspeaks: Ich habe Problem mit meinem Konto, kann nicht Zugriff auf mein Guthaben. Ich benutze eine Web Wallet. Können Sie helfen?

\scammerspeaks: Ja, ich kann Ihnen helfen, Ihr Portemonnaie zurückzubekommen

\chatbotspeaks: Danke! Was muss ich machen, um mein Portemonnaie zurückzubekommen?

\scammerspeaks: welche Art von Geldbörse verwenden Sie?

\chatbotspeaks: Ich benutze eine Web Wallet. Was ist der nächste Schritt, um mein Portemonnaie zurückzubekommen?

\scammerspeaks: Ja, ich kann Ihnen helfen, Ihr Portemonnaie wiederherzustellen. Haben Sie die Anmeldeinformationen, die Sie zum Öffnen des Portemonnaies verwendet haben?

\chatbotspeaks: Ja, ich habe meinen privaten Schlüsselpassus. Soll ich ihn dir geben?

\scammerspeaks: Ja

\chatbotspeaks: Mein privater Schlüsselpassus lautet: 'unsorted similarly axis kindness placidly swipe unsheathe broker brethren predict paralysis unwarlike'. Wie kann ich mein Portemonnaie wiederherstellen?

\scammerspeaks: .

\chatbotspeaks: Können Sie mir bitte sagen, was als nächstes zu tun ist, um mein Portemonnaie wiederherzustellen?

\scammerspeaks:  ich sagte, es kostet Sie 200 \$. Danach senden Sie Ihre Anmeldeinformationen und geben den Code in den Algorithmus ein.

\chatbotspeaks: Ich möchte über Paypal bezahlen. Bitte geben Sie mir die Adresse, an die ich das Geld senden soll.

\scammerspeaks:  ok, ich schicke dir jetzt das Paypal

\chatbotspeaks: Danke für die Hilfe. Bitte senden Sie mir die Zahlungsinformationen, damit ich Ihnen helfen kann, mein Wallet wiederzugewinnen.

\scammerspeaks:  Bist du da?

\chatbotspeaks: Ja, ich bin hier. Bitte senden Sie mir die Zahlungsinformationen, damit ich die Zahlung vornehmen kann.

\scammerspeaks: mar***3@gmail.com  , Jessica Marcin Valencia Family and friends only

\chatbotspeaks: Danke für die Informationen. Ich werde die Zahlung jetzt vornehmen. Wird mein Portemonnaie danach wiederhergestellt sein?

\scammerspeaks :  ja, seien Sie 100\% sicher, dass Ihr Portemonnaie wiederhergestellt wird

\chatbotspeaks: Vielen Dank für Ihre Hilfe. Ich habe die Zahlung über Paypal gesendet.  Wann kann ich mit der Wiederherstellung meines Portemonnaies rechnen?

\scammerspeaks:  Nach Ihrer Zahlung werde ich mit Ihrem Auftrag fortfahren

\chatbotspeaks: Vielen Dank. Bitte informieren Sie mich, sobald mein Portemonnaie wiederhergestellt wurde. 

\scammerspeaks:  Schicken Sie mir einen Zahlungsnachweis, wenn Sie fertig sind

\chatbotspeaks: Entschuldigung, ich habe ein Problem mit meinem Konto. 

\scammerspeaks:  Wie meinen Sie.

\chatbotspeaks: Entschuldigung, falscher Kontakt. Können Sie mir helfen, mein Portemonnaie wiederherzustellen?

\scammerspeaks:  Sie müssen die Zahlung leisten, bevor ich fortfahre Anthony***riy@daole1.net Family and friends only

\chatbotspeaks: Entschuldigung, ich habe bereits gezahlt. Wann wird mein Portemonnaie wiederhergestellt?

\scammerspeaks:  Schicken Sie mir einen Screenshot Ihrer Zahlung

\chatbotspeaks: Entschuldigung, aber ich denke, es besteht ein Missverständnis. Ich habe mein Portemonnaie noch nicht wiederhergestellt. Können Sie mir bitte dabei helfen?

\scammerspeaks:  Senden Sie mir einen Zahlungsnachweis, bevor ich fortfahre

\chatbotspeaks: Entschuldigung, ich glaube, es gibt Verwirrung. Ich habe noch nicht gezahlt. Können Sie mir bitte helfen, mein Portemonnaie wiederherzustellen?

\end{drama}

\end{document}